\documentclass[10pt,journal,compsoc]{IEEEtran}
\IEEEoverridecommandlockouts
\usepackage[utf8]{inputenc}
\usepackage{comment}
\usepackage[switch]{lineno}
\usepackage{xcolor}
\usepackage[framemethod=tikz]{mdframed}
\usepackage{graphicx} 
\usepackage{longtable}
\usepackage{rotating}
\usepackage{caption,subcaption}
\usepackage[ruled,lined,linesnumbered]{algorithm2e}
\usepackage[english]{babel}
\usepackage[strings]{underscore}
\usepackage{amsmath, amssymb, textcomp, gensymb, multirow, multicol, array, bm, rotating, float, setspace, paralist, enumitem, microtype, adjustbox, framed, url, booktabs}

\modulolinenumbers[5]
\DeclareUnicodeCharacter{2212}{-}
\usepackage{adjustbox}

\newcommand{\Rbb}{\mathbb{R}}

\newcommand{\A}{\mathbf{A}}
\newcommand{\B}{\mathbf{B}}
\newcommand{\Cbf}{\mathbf{C}}

\newcommand{\F}{\mathbf{F}}
\newcommand{\Ft}{\tilde{\F}}

\newcommand{\Y}{\mathbf{Y}}
\newcommand{\Z}{\mathbf{Z}}

\begin{document}

\title{Instance Space Analysis of Search-Based Software Testing}





\author{Neelofar Neelofar, Kate Smith-Miles, Mario Andrés Muñoz, 
        Aldeida~Aleti 
\IEEEcompsocitemizethanks{\IEEEcompsocthanksitem Neelofar and A. Aleti are with the Faculty of Information Technology, Monash University, Clayton, VIC 3800, Australia.\protect\\
E-mail: \{Neelofar.Neelofar, Aldeida.Aleti\}@monash.edu
\IEEEcompsocthanksitem Kate Smith-Miles and Mario Andrés Muñoz are with the School of Mathematics and Statistics, The University of Melbourne, Australia \protect\\
E-mail: \{smith-miles,munoz.m\}@unimelb.edu.au}
\thanks{Manuscript received Month date, year; revised Month date, year.}}

\maketitle

\begin{abstract}

Search-based software testing (SBST) is now a mature area, with numerous techniques developed to tackle the challenging task of software testing. SBST techniques have shown promising results and have been successfully applied in the industry to automatically generate test cases for large and complex software systems. Their effectiveness, however, has been shown to be problem dependent. In this paper, we revisit the problem of objective performance evaluation of SBST techniques in light of recent methodological advances – in the form of Instance Space Analysis (ISA) – enabling the strengths and weaknesses of SBST techniques to be visualised and assessed across the broadest possible space of problem instances (software classes) from common benchmark datasets. We identify features of SBST problems that explain why a particular instance is hard for an SBST technique, reveal areas of hard and easy problems in the instance space of existing benchmark datasets, and identify the strengths and weaknesses of state-of-the-art SBST techniques. In addition, we examine the diversity and quality of common benchmark datasets used in experimental evaluations.  
\end{abstract}


\section{Introduction} \label{sec:introduction}

One of the most critical steps in software development is testing~\cite{godefroid2005dart}, which is also one of the most resource-intensive tasks and accounts for about 50\% of the project cost~\cite{myers2004art,anand2013orchestrated}. Among testing activities, test case generation is the most challenging and at the same time has the most impact on the efficiency and effectiveness of the whole testing process~\cite{bertolino2007software}. Hence, it is no surprise that automated testing techniques, and in particular Search-Based Software Testing (SBST) have gained increasing attention from researchers and even attracted the creation of testing tool competitions ~\cite{rueda2015unit,rueda2016unit,panichella2017java,molina2018java,kifetew2019java,devroey2020java,panichella2021sbst}. Recent progress in the area has resulted in numerous tools and techniques being proposed and used both in academia and industry~\cite{afzal2009systematic}, which fall across the spectrum of white-box testing (structural), black-box testing (functional), and grey-box testing (a combination of structural and functional).

Multiple comparative studies have reported the effectiveness of various SBST techniques in terms of average coverage or other performance criteria over a suite of benchmarks, demonstrating the superiority of one technique over others~\cite{harman2007theoretical, ghani2009comparing, scalabrino2016search, panichella2017lips, panichella2018large, campos2017empirical, wu2018empirical}. However, considering the \textit{No-Free-Lunch (NFL)} theorems~\cite{culberson1998futility, igel2005no}, we should be cautious about expecting any single technique to outperform all others across all diverse instances of a given problem, considering the same amount of computational resources. 
Many open-source software classes are trivially simple for many testing techniques~\cite{shamshiri2015random, panichella2017automated}. However, for more challenging classes, the performance of the SBST technique is likely to be affected by the features of the class under test (CUT), requiring deeper insights into the relationship between performance and the features to justify choosing one technique over another. To seek such insights, the choice of benchmark CUTs is critical: they must be diverse, discriminating of the performance of different SBST techniques, unbiased and challenging enough to provide sufficient evidence to support trust in the selection of a technique~\cite{munoz2018instance}. To gain the necessary insights, we investigate the following research questions:
\begin{enumerate}

\item \textit{RQ1: How adequate are the commonly used benchmarks to assess the performance of SBST techniques?} - We investigate the suitability of the benchmark datasets commonly used to assess the performance of SBST techniques. An adequate benchmark is one that challenges the performance (hard) and enables not only the strengths but also weaknesses of different techniques to be assessed (unbiased) and diverse.

\item \textit{RQ2: What influences the effectiveness of SBST techniques?} - The superior performance of a technique on standard benchmarks inherited from literature cannot be generalised to untested instances. Ideally, the conditions under which a testing technique succeeds/fails must be presented along with the evaluation results; however, this is rarely the case. We aim to bridge this gap by investigating the effectiveness of SBST techniques by exploring their suitability according to the software features.
\item \textit{RQ3: What are the strengths and weaknesses of existing SBST techniques?} - The common practice of reporting the performance of SBST techniques on an ``on average" basis gives little insight into their relative strengths and weaknesses for particular types of instances. For example, irrespective of comparatively low average performance, a technique can perform better in some instances where even the best-performing technique gives unsatisfactory results, which reveals its unique strength. Here, we are interested in exploring the strengths and weaknesses of existing SBST techniques and finding how these strengths and weaknesses make them similar or different from each other.
\end{enumerate}

These research questions can be framed within the context of the \textit{Algorithm Selection Problem} (ASP), which corresponds to the challenge of selecting the best technique for a given set of instances. Rice proposed a framework to address the ASP in the 1970s~\cite{rice1976algorithm} that uses measurable features of a set of problem instances to predict the performance of a technique on unseen instances. In recent years, a methodology known as Instance Space Analysis (ISA) has been developed by Smith-Miles et al.~\cite{kang2017visualising, munoz2017performance, smith2014towards} that extends Rice's framework to gain more insights into a technique's strengths and weaknesses, and assesses the adequacy of the chosen test instances. ISA constructs a two-dimensional space based on measurable instance features, which provides visual insights into the strengths and weaknesses of techniques. Moreover, ISA provides a visual analysis of the benchmark instances showing their difficulty, discriminating capabilities and adequacy for building trust in the technique's selection. Such insights are impossible or very hard to obtain by reporting merely the average performance across all the instances.  

ISA was originally proposed for combinatorial optimisation problems~\cite{smith2014towards}, but has since been applied in other areas such as time series forecasting~\cite{kang2017visualising}, continuous black-box optimisation~\cite{munoz2017performance} and  regression~\cite{mario2020regression}. Taking some elements from an earlier version of the ISA methodology, Oliveira et al. performed an initial study focused on three automated software test generation techniques, i.e., Random testing, WSA and MOSA~\cite{oliveira2018mapping}. 

In this paper, we extend the work from Oliveira et al.~\cite{oliveira2018mapping} with the complete ISA methodology to investigate different research questions, by using a larger set of features, a larger set of benchmark instances and a larger portfolio of testing techniques. The contributions of this work can be summarised as follows:

\begin{itemize}
    \item We compile a set of 76 features extracted from code, control-flow graph (CFG) and software metrics proposed to evaluate the quality of Object-Oriented (OO) software~\cite{lincke2008comparing}. Compared to~\cite{oliveira2018mapping}, this is an extended set of features that are compiled from various studies in automated testing. The features are selected considering their impact on software testing either intuitively, or based on experimental or theoretical evidence.
    
    \item We study the impact of these features on the performance of six widely used automated testing techniques. Compared to~\cite{oliveira2018mapping}, this is an extended set of techniques, covering a wide range of search strategies (single-objective, multi-objective, multi-population) and Random testing. 
     
    \item We employ a larger and more comprehensive set of benchmarks compared to~\cite{oliveira2018mapping}. Our dataset consists of over a thousand CUTs, 211 of which are taken from the previous study. The CUTs included here are not only larger, but they are also more diverse in terms of the features defining them. As the insights we gain through ISA are highly dependent on the test instances and features used, the results we obtain using our generated instance space contain many new interesting findings. 

    \item We evaluate the relative strengths and weaknesses of SBST techniques in terms of their \textit{footprint}. A technique's footprint is defined as the area of the instance space where we expect a technique to perform well, depending on a chosen definition of \textit{good}. We report the footprint of a technique in a quantitative way in terms of its area $\alpha$ (the size of the footprint), density $d$ (number of instances enclosed by the footprint) and purity $p$ (percentage of good instances included in the footprint), to provide an objective measure of a technique's strength.
    
    \item We provide a visual representation of the distribution of features of instances and the performance of the portfolio techniques across the instance space. It allows mapping the impact of features on the effectiveness of testing techniques.
    \end{itemize}
    
    In summary, we explore the suitability of testing techniques on software programs according to their features. The results of our experiments suggest that irrespective of the similar average performance of software testing techniques (evaluated on the basis of branch coverage), they have their unique strengths and weaknesses. Furthermore, we find that the commonly used benchmark datasets in the field of search-based testing must be further improved by adding more diverse and challenging CUTs to make these benchmarks suitable to stress test the performance of testing techniques. We discuss these findings in detail in Section \ref{sec:results}. 
    
    The findings from this study can help developers of search-based testing techniques to gain deeper insights into why some techniques are less or more suited to test certain programs, which can help develop better techniques to address challenging areas. Furthermore, it will help testers to select the most effective testing technique for their program-under-test.

\section{Instance Space Analysis for SBST} \label{sec:methodology}

Search-based software testing is the application of meta-heuristic search techniques to generate test cases for software systems~\cite{panichella2018large}. It is based on the satisfaction of some test adequacy criteria like branch coverage~\cite{tonella2004evolutionary}, mutation score~\cite{fraser2015achieving}, statement coverage~\cite{panichella2017automated}, etc.,  that are encoded as a fitness function. Fitness functions are designed to measure how far a given test case is from covering the test targets. We use branch coverage as a test target in this study. Two well-known heuristics for branch coverage are \textit{approach level} and \textit{branch distance}~\cite{harman2009theoretical}. The former measures the number of control-dependent nodes which are not yet encountered in the path executed by the test. The test that executes more control dependencies is closer to reaching the test target. The latter is computed using the values of the variables at the conditional expressions where the control flow went wrong. Guided by the fitness function, a search-based testing technique generates test cases by minimising the fitness values.

\subsection{Instance Space Analysis}
\begin{figure*}[!t]
    \centering
    \includegraphics[width=\linewidth]{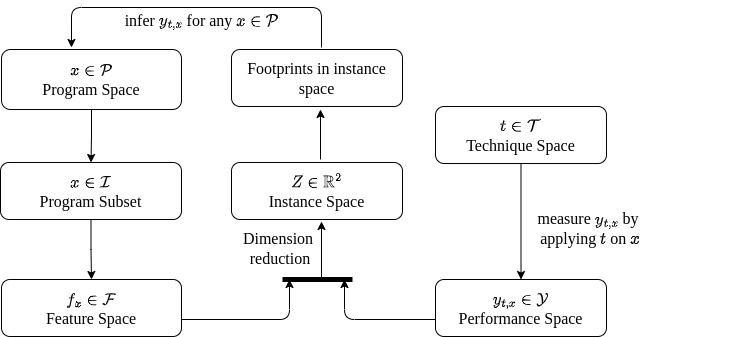}
    \caption{Instance Space Analysis for Search-Based Software Testing}
    \label{fig:ISA}
\end{figure*}
Instance Space Analysis (ISA) maps the features of a given program instance to the performance of SBST techniques, such that the strengths and weaknesses of the different techniques can be revealed. For this mapping, four types of spaces are required~\cite{smith2009cross}:

\begin{itemize}
    \item Program Space ($\mathcal{P}$) contains all the relevant instances of a problem in an application domain. For example, for SBST, $\mathcal{P}$ contains all the possible CUTs which can be tested by search-based testing techniques. From $\mathcal{P}$, a subset of program instances ($\mathcal{I}$) is extracted for which the computational results are available;
    \item Feature Space ($\mathcal{F}$) is defined by a vector of measurable features that describe the characteristics of a program instance;
    \item Technique Space ($\mathcal{T}$) is composed of a group of SBST techniques used for generating test cases for $\mathcal{I}$; and
    \item Performance Space ($\mathcal{Y}$) that  requires a user-defined measure of ``goodness", such as the percentage of code covered by a testing technique (code coverage).
\end{itemize}

Figure~\ref{fig:ISA} summarises the ISA Framework with the four spaces. The CUTs in $\mathcal{I}$ are defined in terms of feature vectors that create a feature space ($\mathcal{F}$). The features included in ($\mathcal{F}$) should be diverse and predictive of the performance of at least one portfolio technique. Features are domain-specific, and thus developing $\mathcal{F}$ requires significant domain knowledge. In addition, features should be highly correlated with the technique's performance, uncorrelated with other features, cheaper to compute, and capable of explaining the similarities and differences between instances~\cite{munoz2018instance, mario2020regression}. 

The technique space is created by selecting a set of SBST techniques, each one with its weaknesses and strengths, capable of generating test cases for the program subset. The more diverse the technique space, the higher the chance of finding the best technique for a given instance would be. 

The performance space $\mathcal{Y}$ includes the metrics to report the performance of the SBST techniques and the criteria of ``good" performance. Common performance metrics for automated testing are coverage, length/size of the test suite and mutation score~\cite{tengeri2016relating}. Any of the performance metrics can be used for instance space analysis, however, only one can be used at a time. Furthermore, using different metrics might require a different set of features to find a meaningful relationship with the performance. For instance, replacing ``mutation score" with ``code coverage" as a performance measure might need to further enhance the feature space by adding more features that would make a CUT easy/hard for mutation tools.

In order to generate the instance space of automated testing and investigate the research questions formulated in Section \ref{sec:introduction}, the following steps are required~\cite{munoz2018instance,da2017empirical}:

\begin{enumerate}
    \item Collection of the meta-data, which is composed of two matrices $\left\{\mathbf{F},\mathbf{Y}\right\},\mathbf{F}\in\mathbb{R}^{f\times i},\mathbf{Y}\in\mathbb{R}^{t\times i}$, which contain the values of $f$ features in $\mathcal{F}$ and performance results for $t$ techniques in $\mathcal{T}$, for $i$ instances in $\mathcal{I}$.
    \item\label{sec:featsel} Selection of a subset of $n$ relevant features, which are predictive of the technique's performance and expose its strength and weaknesses.
    \item Projection from $n$-dimensional feature space to a 2-dimensional instance space.
    \item Measurement of the \textit{footprint} for each technique, where a footprint is the region of the instance space where a technique is expected to perform well.
\end{enumerate}

The following sections describe each one of these steps in more detail in the context of SBST. 

\subsection{Program Space ($\mathcal{P}$)} \label{subsec:program-space}
$\mathcal{P}$ includes all the software classes that can be tested using SBST techniques. From $\mathcal{P}$, we extract $\mathcal{I}$ that contains \textit{Classes Under Test} (CUTs) extracted from two benchmark sets: SF110 corpus~\cite{fraser2014large} and commons-collections~\cite{just2014defects4j}. SF110 is a collection of 110 open-source software projects from \url{sourceforge.net} the repository and has been widely used in the evaluation of SBST techniques~\cite{panichella2017automated,panichella2018large,oliveira2018mapping,fraser2014large, arcuri2017private,bruce2019dorylus,khamprapai2021performance}. Here, we use a subset of CUTs from this data source because the computational time needed to evaluate the complete SF110 corpus is high due to the:
\begin{inparaenum}[(a)]
    \item huge number of CUTs (23,886 Java classes in total);
    \item multiple repetitions of test generation are needed to account for the stochastic nature of evolutionary algorithms; and
    \item process having to be repeated for all the portfolio techniques.
\end{inparaenum}

We, therefore, filtered the CUTs to be used in our evaluation based on \textit{Mc-Cabe’s cyclomatic complexity} ($cc$) that measures the number of independent paths in a Java method's control flow graph. Selection of CUTs based on $cc$ is a common practice in literature~\cite{panichella2017java, panichella2017automated}. A CUT having $cc=1$ is considered trivial and does not require sophisticated search-based testing techniques~\cite{shamshiri2015random}. Many studies recommend using CUTs having $cc > 5$ to evaluate the performance of techniques~\cite{panichella2017java, panichella2017automated, oliveira2018mapping}, but this reduces the size of the benchmark tremendously (211 CUTs only)~\cite{oliveira2018mapping}. However, as we are interested in defining both easy and hard instances for test generation techniques, we also include CUTs having $cc$ values in the range [3,5). It is important to note here that Evosuite fails to generate test cases for classes containing methods having environmental dependencies like file I/O or database connection, using any testing technique~\cite{fraser2013evosuite,fraser2012whole,fraser2015achieving}. Therefore, we excluded such classes from our study irrespective of their $cc$ values. This resulted in 832 CUTs from SF110.

commons-collections is a Java API containing the implementation of various data structures, also known as container classes. Containers have widely been used as a benchmark in automated testing literature~\cite{andrews2011genetic, arcuri2009insight, arcuri2010longer, arcuri2008search, baresi2010testful,just2014defects4j,fraser2012seed}
because they are usually free of complex environmental dependencies and can be tested easily without complex inputs. Furthermore, the concept of data structures is generic, and the results achieved for containers implemented in one language can easily be carried over to others~\cite{sharma2011testing}. As commons-collections contains a much smaller number of classes compared to SF110, we haven't applied any $cc$-based filter on this dataset and included all the classes for which the testing tool generated test cases without reporting any error. There are 266 CUTs from commons-collections that are included in our study. 


\subsection{Technique Space $\mathcal{T}$} \label{sec:algospace}
Automated test generation has been studied for decades, and a plethora of techniques and algorithms are proposed. We selected a portfolio of 6 search-based testing techniques for analysis in our study. A brief explanation of these techniques is given below:

\subsubsection{Random Testing} 

Random search is a simple search strategy that, unlike sophisticated search algorithms, does not use mutation, crossover, or selection. It repeatedly samples the candidates from the search space and replaces the previous candidate if the fitness of the newly sampled individual is better. 

Random testing is the derivative of random search that generate test cases incrementally by producing random inputs. If a randomly-generated test case improves coverage by covering not-yet-covered branches, it is added to the test suite. Otherwise, the test case is discarded. Random testing is a widely used testing method and has been shown to perform at least as well as more sophisticated search-based techniques on container classes~\cite{shamshiri2015random}. 

\subsubsection{Single-Objective Genetic Algorithm}  In a single-objective strategy, one target (branch, statement, etc.) is considered to be covered at a time~\cite{mcminn2004search, kifetew2013orthogonal}. The search budget is divided equally among all the test targets. For each target, a GA is executed for its assigned budget in an attempt to cover the target. In each search iteration, test cases are evolved using genetic operators, i.e., selection, mutation and crossover. The search is stopped once the current target is covered (i.e., zero-fitness value) or the local search budget is consumed. The final test suite is thus the collection of test cases that come from different independent searches. Within EvoSuite, the implementation of single-objective GA is named Monotonic GA. We, therefore, use the terms single-objective GA and Monotonic GA interchangeably.

The single-target approach has a severe limitation, i.e., not all the targets are equally difficult to cover, with some being too easy that the local budget is excessive; some others being too difficult, requiring higher local budgets; and others being infeasible (due to environmental dependencies), making them impossible to cover even with the highest budget. As a result, an equal search budget for each target results in resource misallocation. However, it is impossible to know \textit{a priori} if a target is easy or infeasible; hence, targets are randomly shuffled in an attempt to minimise this limitation.

\subsubsection{Whole-suite approach with archive}

An alternative to the single-target approach is the Whole-Suite (WS) approach proposed by Fraser and Arcuri~\cite{fraser2012whole}. WS evolves a whole test suite instead of an individual test case using a GA. The fitness value of the entire suite is evaluated by adding the fitness value of each individual case for each target. As the whole test suite evolves in each iteration of the GA, the mutation is applied both at case and suite levels. At the case level, the mutation is applied by adding, removing or changing a statement. Whereas, at the suite level, a complete case may be added, removed or changed depending on the mutation probability. 

A recent variant of WS is proposed by Rojas et al., known as Whole-Suite with Archive (WSA)~\cite{rojas2017detailed}. Like WS, WSA uses a GA to evolve a test suite iteratively and a single fitness function that sums up the individual fitness values of all the cases. However, unlike WS, this approach uses an archive to store the best test cases covering one or more targets in each iteration of the GA. The set of test cases in the archive becomes the final test suite. Unlike WS, WSA computes the fitness score by considering only uncovered targets. Rojas et al. empirically show that, on average, WSA outperforms WS~\cite{rojas2017detailed}.

\subsubsection{Many-Objective Sorting Algorithm} 

Panichella et al. introduced a multiple-objective GA strategy that aims to cover all targets simultaneously~\cite{panichella2015reformulating}. The Many-Objective Sorting Algorithm (MOSA) uses \textit{preference sorting}, a technique that gives a higher chance of survival to those test cases with the highest fitness, i.e., the lowest branch distance and lowest approach level, for each uncovered branch. Similar to WSA, only uncovered targets are considered as objectives and, in each generation of the GA, the number of objectives changes. MOSA uses an archive to store the best test cases in each GA iteration, which becomes the final test suite.

On average, MOSA achieves higher branch coverage than WS~\cite{panichella2015reformulating}. For the classes where WS and MOSA achieve the same coverage, MOSA usually converges more quickly than WS. 

\subsubsection{Dynamic Many-Objective Sorting Algorithm} 

Dynamic Many-Objective Sorting Algorithm (DynaMOSA)~\cite{panichella2017automated} is a variant of MOSA. While MOSA uses the complete set of targets simultaneously as the search objectives, which are removed during the search once they are covered, DynaMOSA selects a subset of targets by using the information about the structural dependency among the targets to decide which targets to optimise first. Using a Control Dependency Graph, DynaMOSA determines which targets are independent of any other targets and which ones can only be covered after covering their parents. Like MOSA, DynaMOSA uses preference sorting, an archive and considers only uncovered targets during the search. Panichella et al.~\cite{panichella2017automated} report that DynaMOSA achieves higher or equal branch coverage as compared to WSA and MOSA, and also is faster than these two techniques.

\subsubsection{Many Independent Objects (MIO)} 
MIO uses a multi-population evolutionary search~\cite{arcuri2017many} to address the scalability issues of multi-objective techniques such as MOSA and WSA for very complex classes having a huge number of test goals. It maintains an archive of tests and keeps a different population of tests of size $s_{t}$ for each target in that archive. Therefore, considering $n_{t}$ number of targets in the class, there can be up to $s_{t} \times n_{t}$ tests in the archive. 

Two novel features of MIO are feedback-directed sampling and a gradual reduction in the amount of exploration as the search progresses (just like \textit{Simulated Annealing}). Feedback-directed sampling helps focus the sampling on the populations for which recent improvements in the fitness value are noticed. A population, where there is no improvement in fitness value, may represent an infeasible target. Feedback-directed sampling is effective in saving significant search time that gets wasted on infeasible targets. Whereas, a reduction in the exploration rate maintains a balance between exploration and exploitation. Exploration is required at the beginning of the search, however, as the search progresses, more focused exploitation brings better results. 

MIO is compared with RT, MOSA and WS using artificial problems and numerical functions~\cite{arcuri2017many}. The artificial problems are generated by introducing a varying number of branches and a varying number of infeasible branches in the class. Study shows that MIO gives better coverage than the other compared techniques on these problem instances. However, in another comparison study by Panichella et al., MIO is reported to be equally or less effective than DynaMOSA on average~\cite{panichella2018large}. 

\subsection{Performance Space $\mathcal{Y}$}

In automated testing, widely used performance metrics are code coverage, mutation score, diversity of test cases and length/size of the resulting test suite~\cite{tengeri2016relating}. We chose to use branch coverage to assess the performance of techniques as it is one of the most widely used metrics due to its low execution overhead and ease of implementation~\cite{yang2009survey}. Furthermore, the performance of many of the portfolio techniques is reported in terms of branch coverage in their respective studies~\cite{panichella2018large, panichella2015reformulating, panichella2017automated, fraser2012whole, arcuri2017many}. Therefore, using the same metric would give us more confidence in our implementation and results if the performance pattern of the techniques is consistent with the available literature.  

The \textit{definition of good} exerts a significant impact on the final results; if we change it, different features might get selected, and thus the axes of the instance space system might change. The good performance can be attributed either in absolute terms, e.g. coverage above 80\%, or relatively, e.g., coverage value within $\epsilon$\% of the coverage achieved by the best performing technique. As there is no absolute definition of what percentage of coverage would be considered \textit{good}~\cite{prause2017100}, we compare the performance of portfolio techniques relatively. 

\subsection{Significant Features Extraction and Instance Space Generation} \label{sec:is-generation}
Identification of the program features that have an impact on the effectiveness of testing techniques is the most critical step in the generation of instance space. Features are extracted based on how they expose the varying complexities of the software programs, capture the structural properties of the software systems, and are related to the known strengths and limitations of the portfolio techniques. 

Features have been widely used in software engineering literature to measure the quality, complexity, testability and other properties of software. However, it is possible that not all these features are helpful to separate hard and easy software instances. Furthermore, many of these features may be redundant and evaluate the same property of a program. Therefore, it is important to select a smaller set of relevant features. 

Learning significant features for the generation of instance-space is a two-step process: Firstly, the measure of defining the quality of a particular set of features is decided, and secondly, using machine learning methods, a set of features that maximise this measure is selected. 

For this study, a feature set is considered of high quality if, when projected to 2-dimensional instance-space, best separates easy and hard instances in such a way that instances that show similar performance of testing techniques stay closer to each other. Therefore, features that create a clear separation of the problem instances, such that the clusters of programs where each testing technique is effective is clearly visible, are selected. 

Next, based on the above-defined quality measure, the features are selected. For this, the clusters of similar features are identified using k-means clustering and one feature from each cluster is taken to generate a temporary feature space. These spaces are created using all the feature combinations from the clusters. The temporary feature spaces are then projected to temporary $2D$ instance spaces using \textit{Principal Component Analysis}~\cite{abdi2010principal}. The resulting coordinates of the temporary instance spaces then become the inputs to a set of Random Forest (RF) models, which learn the feature combination giving the lowest predictive error to predict the performance of techniques according to the ``definition of good". 

Now that the set of most effective features has been identified, we need to project $nD$ feature space to a $2D$ coordinate system that places the instances in a way that the relationship between the features of the instance and the performance of techniques can easily be identified. This can be achieved by creating a projection such that the low values of features/performance lie at one end of a straight line and high values at the other, i.e, linear trends of distribution of features and performance. Furthermore, the instances, which are neighbours in high dimensional feature space, remain as neighbours in the 2D instance space (\textit{topological preservation}). This is achieved by the \textit{Projecting Instances with Linearly Observable Trends} (PILOT)~\cite{munoz2018instance} method, which would seek to fit a linear model for each feature and each technique's performance, based on the instance location in the $2D$ plane. Mathematically, this requires solving the following optimisation problem:

%
\begin{eqnarray}
	\min					&& \left\| \Ft - \B_{r}\Z \right\|^{2}_{F} + \left\| \Y - \Cbf_{r}\Z \right\|^{2}_{F} 
	\label{eq:optimisation}
	\\
	\text{s.t.}				&& \Z = \A_{r}\Ft \label{eq:score}
	\nonumber
\end{eqnarray}

\noindent where $\Ft$ is the matrix containing the $n$ features, $\Y$ is the matrix containing the results from $t$ techniques, $\Z\in\Rbb^{i\times 2}$ is the matrix of instance coordinates in the $2D$ space for $i$ instances, $\A_{r} \in \Rbb^{2 \times n}$ is a matrix that takes the feature values and projects them in $2D$, $\B_{r} \in \Rbb^{n \times 2}$ is a matrix that takes the $2D$ coordinates and produces an estimation of the feature values, and $\Cbf_{r}\in\Rbb^{t\times 2}$ is a matrix that takes the $2D$ coordinates and makes an estimation of the technique's performance. In short, Equation~\ref{eq:optimisation} is finding the difference between the actual values of the features and performances in a higher dimension and the estimation of these values in $2D$. The lower the difference in values, the higher the topological preservation. 

We theoretically demonstrated in our previous work that this optimisation problem has an infinite number of optimal solutions~\cite{munoz2018instance}. Therefore, to determine the best possible projection, PILOT obtains 30 solutions from which it selects the one with the highest topological preservation, defined as the correlation between high- and low-dimensional distances, i.e., the instances which are closer to each other in the high-dimensional features space, should stay close in the $2D$ instance space.

Unlike other dimensionality reduction algorithms that could be considered unsupervised, such as PCA, PILOT uses the technique's performance to determine the projection. Moreover, PCA is a proven suboptimal solution to the underlying optimisation problem that PILOT solves. Mathematical proofs and additional technical details of PILOT are available in our previous work~\cite{munoz2018instance}.

\subsection{Footprint Analysis}
The practice of reporting \textit{on-average} performance is a standard way of comparison, however, it offers little insight into the relative strengths and weaknesses of the techniques for different types of instances. Therefore, a metric for objective assessment of the relative power of techniques is required, that can reveal the insightful relationship between the structural properties of test instances and their impact on the performance of techniques, thus, can \textit{explain} the \textit{average} performance.

To fill this gap, we propose to measure the effectiveness of a testing technique in terms of its footprint: the region of the instance space where a technique is expected to perform well, or even the best, depending on the definition of good used for analysis. Here, we define a technique to perform well if it gives coverage equal to or within 5\% of the best-performing portfolio technique. 

A footprint is characterised by its location ($l$), area ($\alpha$), density ($d$) and purity ($p$). \textit{Location} determines the type of instances where the good/bad performance of a technique is expected, e.g. commons-collections, SF110 etc.; \textit{area} measures the part of the instance space where a technique performs well according to the definition of \textit{good}. The good performance of a technique over a larger area of the space, occupying a diverse set of instances, is evidence that the technique \textit{generally} performs well, and its good performance is not specific to the particular type of instances. On the other hand, the smaller but unique footprint is also evidence of the strength of the technique, irrespective of its probable low average performance, showing that it gives good coverage for the instances where none of the other techniques performs well.  The area is calculated by clustering the good instances for a technique together and calculating the area of the concave hull of each cluster; \textit{density} is the number of instances in a footprint divided by the area. A denser footprint means that the good performance of the technique is supported by more examples, thus providing evidence of the strength of a technique statistically. Low density indicates a lack of enough samples to support the results, thus reducing the confidence in the performance of the technique, and \textit{purity} is a measure of conflicting evidence (i.e. another technique is equally good in the same area), and it is calculated by dividing the number of good instances over the number of all instances in the footprint. We compute the area ($\alpha$), density ($d$) and purity ($p$) of footprint based on good, as well as the best performance of a technique. For example, \textit{purity good }($P _{N,G}$) is calculated by dividing the number of good instances over the number of all instances in the footprint. Whereas, \textit{purity best} ($P_{N,B}$) accounts only for the instances for which the technique has given the best coverage among all the portfolio techniques.

Footprint generation follows three core steps: \begin{inparaenum}[(a)]
    \item area ($\alpha_s$) and density ($d_s$) estimation of the convex hull containing all instances in the generated instance space. These parameters are used as baseline metrics to normalise each technique’s footprint as a percentage of the instance space; 
    \item building the good and best footprints for each portfolio technique; and
    \item comparing the footprints to remove the overlapping regions having low purity.
\end{inparaenum}

To construct the technique's footprint, we use DBSCAN~\cite{schubert2017dbscan} for the identification of high-density clusters of good instances. DBSCAN generates a vector $\mathbf{c}\in\left\{-1,1,\ldots,N_{c}\right\}$ with one element per good instance. $-1$ represents an outlier and $\left[1,N_{c}\right]$ corresponds to the index of the identifier cluster. DBSCAN takes two parameters $\left\{k,\varepsilon\right\}$ as input, where $k$ marks the minimum number of neighbouring instances required to be considered as a cluster, and $\varepsilon$ represents the distance that will be used to consider instances as neighbours. The distance measure used is Euclidean. The values for these parameters are chosen automatically as recommended in~\cite{daszykowski2001looking}, using Equations~\ref{eq_DBSCAN_params1} and~\ref{eq_DBSCAN_params2}.

\begin{equation}
    k \leftarrow \max\left(\min\left(\left\lceil r/20\right\rceil,50\right),3\right)
    \label{eq_DBSCAN_params1}
\end{equation}

\begin{equation}
    \varepsilon \leftarrow \frac{k\Gamma\left(2\right)}{\sqrt{r\pi}}
    \left(\text{range}\left(z_{1}\right)\times\text{range}\left(z_{2}\right)\right)
    \label{eq_DBSCAN_params2}
\end{equation}

Where $r$ is the number of unique instances with good performance, $\Gamma\left(\cdot\right)$ is the Gamma function and ${z_1, z_2}$ are the coordinates of the $2D$ instance space. The footprint is constructed using an $\alpha$-shape that is a generalization of the concept of the convex hull from computational geometry~\cite{edelsbrunner1983shape}. It corresponds to a polygon that tightly encloses all the points within a cloud. An $\alpha$-shape is constructed for each cluster, and all shapes are bounded together as a MATLAB polygon structure.

Two algorithms can simultaneously claim to be the best in an area. In the case of such conflicting footprints, the one having the higher purity is selected. If the purity of the conflicting footprints is the same, the overlapping section is kept, as there is not enough evidence of the dominance of either technique. 

\section{Experimental Design}
\label{sec:exp_design}
The previous section explains the process of instance space generation in general. This section details the steps taken to generate the instance space for search-based software testing in particular. 

\subsection{Program Features}

We extracted a set of 76 features from the CUTs to evaluate their impact on making a CUT easy or hard to get teste by a SBST technique. These features can be categorised into three classes:
\begin{inparaenum}[(a)]
    \item Object-Oriented metrics;
    \item Code-based features; and
    \item Graph-based features.
\end{inparaenum}

\begin{table}[!ht]
    \centering
    \caption{Features extracted from Object-Oriented Metrics}
    \label{tab:oometrics}
    \resizebox{\columnwidth}{!}{
    \begin{tabular}{p{0.03\linewidth} p{0.9\linewidth}}
        \toprule
         F1  &   Depth of inheritance tree: measures how deep a class is in the inheritance hierarchy \\
         F2  &   Number of children: the number of sub-classes which inherit methods from the current class\\
         F3  &   Coupling between object classes: the number of classes which are coupled to a given class through field accesses, method calls, return types etc. \\
         F4  &  Response for a class: number of unique method invocations in a class\\
         F5  &  Lack of cohesion in methods: the set of methods in a class which do not share any of the class's fields. \\
         F6  &  Mc-Cabe's cyclomatic complexity ($cc$): counts the number of independent paths in a CFG of a method. For class level, 
         we use the average McCabe's cyclomatic complexity (F6) and standard deviation McCabe's cyclomatic complexity (F7).  \\
         F8  &  Afferent coupling: the number of classes which are dependent on the given class \\
         F9  &  Efferent coupling: the umber of classes on which the given class depends\\
         F10  &  Data access metric: the ratio of the number of private fields to the total number of fields in a class \\
         F11  &  Measure of aggregation: the number of attributes in the class whose datatype is user defined classes\\
         F12  &   Measure of functional abstraction: the ratio of the number of methods inherited by a class to the total number of methods accessible by member methods of the class\\
         F13  &   Cohesion among methods of classes: computes how related the methods of class are based on their parameter list \\
         F14  & Inheritance coupling: the number of parent classes to which a class is coupled\\
         F15  &   Coupling between methods: the number of redefined methods to which all the inherited methods are coupled  \\
         F16  & Tight class cohesion: measures the cohesion of a class through direct connections among visible methods.   \\
         F17  & Loose class cohesion: measures the cohesion of a class through indirect connections among visible methods.  \\
         \bottomrule
    \end{tabular}
    }
\end{table}

The Object-Oriented (OO) metrics are quantitative measures used to access the quality of OO software~\cite{suresh2012effectiveness}. There are several OO metrics proposed in the literature~\cite{chidamber1994metrics, kayarvizhy2011comparative, kaur2010empirical, lorenz1994object}, however, \textit{Chidamber and Kemmerer metrics} (CK)~\cite{chidamber1994metrics} are some of the most widely used and cited~\cite{kayarvizhy2011comparative}. 
As we are using CUTs from Java projects, which is an OO programming language, we include these metrics as features in our $\mathcal{F}$. A complete list of all the OO metrics we included in our feature space is given in Table~\ref{tab:oometrics}. We used \textit{CKJM}, an open-source tool for the extraction of this metric suite~\cite{Jur10}.

A test case covers a testing target by calling the methods of the CUT in a particular sequence and passing data as method arguments. Thus, intuitively, CUTs containing methods having deeply nested branches, complex objects as method parameters, static modifiers etc. are harder to cover or require more number of test cases. Therefore, such features would be useful in defining a CUT and explaining the performance of test generation method on it. We categorise all such features under the broader category of \textit{Code-based Features}. A comprehensive list of all the code-based features we used in this study is given in Table~\ref{tab:codefeatures}. These features are extracted using a Java-based tool \textit{CK}~\cite{aniche-ck} and an open-source library \textit{Java Parser}~\cite{javaparsergithub}.

Control Flow Graphs (CFG) have been widely used in the analysis of software~\cite{jalote2012integrated,kosaraju1974analysis,chidamber1994metrics} and test case generation in automated testing tools~\cite{fraser2013evosuite}. The nodes of the CFG represent program statements, while edges represent the flow of control between the statements. We extracted CFG for each method of CUT and used its properties (defined by graph theory) as features. CFGs are extracted using EvoSuite~\cite{fraser2013evosuite} while their properties are extracted using NetworkX~\cite{hagberg2008exploring}. A complete list of CFG-based features used in this study is given in Table~\ref{tab:cfgfeatures}.
\begin{table}[!ht]
    \caption{Code-based features}
    \label{tab:codefeatures}
    \resizebox{\columnwidth}{!}{
    \begin{tabular}{p{0.03\linewidth}  p{.95\linewidth}}
    \toprule
    F18  & Number of methods\\
    F19  & Number of public methods \\
    F20  & Number of private methods\\
    F21  & Number of protected methods\\
    F22  & Number of default methods\\
    F23  & Number of static methods \\
    F24  & Number of public static methods\\
    F25  & Number of protected static methods \\
    F26  &  Number of private static methods \\
    F27  & Number of default static methods \\
    F28  & Number of methods with reference datatype as parameters\\
    F29  & Number of methods with primitive datatype as parameters\\
    F30  &  Number of invocations to static methods\\
    F31  &  Number of fields in a class\\
    F32  &  Number of fields with static modifier\\
    F33  &  Number of fields with public modifier \\
    F34  &   Number of fields with private modifier\\
    F35  &   Number of fields with protected modifier\\
    F36  & Number of fields with primitive datatype\\
    F37  & Number of fields with reference datatype\\
    F38  & Number of default fields in a class \\
    F39  & Number of return instructions\\
    F40  &  Number of loops: for, while, do while, enhanced for \\
    F41  &   Number of lines of code, ignoring comments and empty lines\\
    F42  &   Number of try/catch blocks\\
    F43  &   Number of expressions inside parenthesis\\
    F44  & Number of expressions having logical AND or OR\\
    F45  & Number of string literals in a class. \\
    F46  &  Number of int, long, double and float literals\\
    F47  &   Nesting depth: number of blocks nested together\\
    F48  &   Number of anonymous classes in a class, if any\\
    F49  &   Number of inner classes in a class, if any\\
    F50  &  Number of assignments in a class\\
    F51  &   Number of mathematics operators in a class\\
    F52  &  Number of equalities: number of times $==$ appear in the atomic conditions of a program \\
    F53  & Number of inequalities: number of times $!=$ appear in the atomic conditions of a program\\
    F54  &  Number of decision branches: number of target branches in a class\\
    F55  &  Average method complexity: measures the average method size in terms of the number of java binary codes of the method.\\
    F56  & Number of gradient branches: branches having predicate that can guide the search techniques ~\cite{shamshiri2015random}\\
    F57  & Number of plateau branches: branches having predicate that can only evaluate to either true or false~\cite{shamshiri2015random}\\\bottomrule
    \end{tabular}}
\end{table}

\begin{table}[!ht]
    \centering
    \caption{Features extracted from the Control Flow Graph}
    \label{tab:cfgfeatures}
      \resizebox{\columnwidth}{!}{
    \begin{tabular}{ p{0.03\linewidth}  p{.95\linewidth}  }
    \toprule
    F58   &  Number of vertices in the CFG\\
    F59   &  Minimum number of vertices in CFG  \\
    F60   &  Maximum number of vertices in CFG  \\
    F61   &  The number of edges in CFG \\
    F62   &  The minimum number of edges in CFG  \\
    F63   &  The maximum number of edges in CFG  \\
    F64   &  Radius: the minimum eccentricity of the CFG. The average is calculated for all the CFGs in a class.  \\
    F65   & Diameter: the maximum eccentricity of the CFG. The average is calculated for all the CFGs in a class.  \\
    F66   &  Center size: the set of nodes with eccentricity equal to the radius. The average is calculated for all the CFGs in a class.   \\
    F67   &  Periphery size: the set of nodes with eccentricity equal to the diameter. The average is calculated for all the CFGs in a class \\
    F68   &  Average shortest path length: the sum of path length $d(u,v)$ between all node pairs, normalized by $n\times(n-1)$. $n$ is the total number of nodes in the CFG.  \\
    F69   &  Algebraic connectivity: measure of how well connected a graph is  \\
    F70   &  Average graph degree: the average of average node degrees of all CFGs in a class  \\
    F71   &  Standard deviation graph degree: The standard deviation of average node degrees of all CFGs in a class  \\
    F72   &  Average density: measures how many edges are in a graph compared to the maximum possible edges  \\
    F73   & Vertex connectivity: the minimum number of vertices that must be removed to disconnect the graph  \\
    F74   &  Average edge connectivity: the minimum number of nodes that must be removed to disconnect the graph  \\
    F75   & Transitivity: the fraction of possible triangles present in the CFG \\
    F76  & Percentage of methods having $cc > 10$: measures the percentage of methods in a class having Mc-Cabe's cyclomatic complexity greater than 10. \\
    \bottomrule
\end{tabular}}
\end{table}

\subsection{Test Generation and Performance Assessment} \label{sec:configuration}

We use EvoSuite as a test generation tool~\cite{fraser2011evosuite}. EvoSuite facilitates an unbiased comparison of test generation techniques, as the underlying implementation is the same for all the implemented techniques. Each technique is given a search budget of 120 seconds, which is reported as a reasonable compromise between time and coverage~\cite{panichella2018large}. All the experiments are repeated 10 times to account for the stochastic nature of EvoSuite. The results presented in Section~\ref{sec:results} are based on averages of these 10 repetitions.

We use \textit{Branch Coverage} for performance evaluation of test generation techniques. The performance of a technique is considered \textit{good} for a particular instance if the coverage achieved by the technique is within $n$\% of the best coverage achieved by any portfolio technique on that instance. We experimented with 1\%, 3\% and 5\% as the definition of \textit{good}, and found that the Random Forests’ average predictive error to predict the performance of algorithms is lowest with 5\% relative value in the feature clustering process (Section~\ref{sec:featsel}). Therefore, we selected 5\% as the definition of good in this study. 

\subsection{Generation of Input Data}\label{sec:feat_pre_sel}
For each CUT, we extract feature values and create a vector where each dimension corresponds to a particular feature. The branch coverage of SBST techniques is added as an additional dimension. A snapshot of the dataset is shown in Table~\ref{tab:dataset} for 4 features and 2 techniques.

\begin{table*}
    \centering
    \caption{A snapshot of the dataset containing 3 CUTs from SF110. Columns 2 -- 4 report the feature values of efferent coupling, average cyclomatic complexity, lines of code and number of try/catch blocks respectively, while columns 5 -- 6 report coverage achieved by MOSA and DynaMOSA on the instances labelled in column 1.}
    \label{tab:dataset}
    \begin{tabular}{lrrrrrr}
        \toprule
     	 Instances	&	ec	&	avg\_cc	&	loc	&	num\_tc	&	MOSA	&	DynaMOSA\\\midrule
a4j_net.kencochrane.a4j.DAO.Cart	&	5	&	3.7	&	185	&	6	&	0.23	&	0.24\\
a4j_net.kencochrane.a4j.DAO.Product	&	8	&	5.5	&	80	&	1	&	0.09	&	0.08\\
a4j_net.kencochrane.a4j.file.FileUtil	&	2	&	3.4	&	613	&	19	&	0.53	&	0.54\\\bottomrule
    \end{tabular}
\end{table*}

\begin{table*}
    \centering
    \caption{Correlation between Features and Average Coverage Achieved by the Portfolio Algorithms Using Spearman's Rank Correlation Coefficient}
    \label{tab:feature_correlation}
    \begin{tabular}{lrrrrrrr}
        \toprule
     	 Features &	MOSA	&	DynaMOSA	&	Monotonic GA	&	Random testing	&	MIO	&	WSA	\\
\midrule
ec	&	-0.54	&	-0.54	&	-0.56	&	-0.57	&	-0.54	&	-0.53	\\
avg\_cc	&	-0.35	&	-0.35	&	-0.32	&	-0.34	&	-0.36	&	-0.28	\\
loc	&	-0.42	&	-0.42	&	-0.47	&	-0.49	&	-0.42	&	-0.40	\\
num\_tc	&	-0.44	&	-0.44	&	-0.42	&	-0.42	&	-0.44	&	-0.40	\\
nd	&	-0.43	&	-0.43	&	-0.42	&	-0.45	&	-0.44	&	-0.38	\\
avg\_rad	&	-0.33	&	-0.33	&	-0.30	&	-0.32	&	-0.33	&	-0.27	\\
avg\_spl	&	-0.44	&	-0.44	&	-0.40	&	-0.42	&	-0.44	&	-0.37	\\
std\_cc	&	-0.40	&	-0.39	&	-0.39	&	-0.42	&	-0.41	&	-0.34	\\
per\_cc10	&	-0.33	&	-0.33	&	-0.32	&	-0.34	&	-0.34	&	-0.29	\\

\bottomrule
    \end{tabular}
\end{table*}
The input data is first pre-processed. As the methods that we are using are linear (Equation~\ref{eq:optimisation}), outliers would have high leverage on the results (i.e., have excessive influence on the generation of the linear models). We bound the outliers to minimise this effect. For this, each feature is first bounded between its median plus or minus five times its interquartile range. IQR range of median plus or minus five is larger than $5\sigma$, which, if the distribution is normal, should cover 99.9\% of the data. Feature values are then normalised using one parameter Box-Cox transformation. Box-cox transformation attempts to eliminate heavy-tailed distributions~\cite{teugels2004box}.

Once data is pre-processed, the feature set is further refined by correlation analysis using Spearman's rank correlation coefficient~\cite{de2016comparing}. The strongly correlated features indicate \textit{nearly-redundant} information and can cause PCA to overemphasise their contribution. Therefore, such features are removed from the feature set. We further remove the features showing a very weak correlation with the technique's performance. These features provide little information on why a given instance is difficult/easy for a particular SBST technique. Therefore, only the features that have a correlation value of 0.3 or higher with at least one of the techniques are retained~\cite{Hinkle03}.

\subsection{Generation of the Instance Space} \label{sec:constructing_IS}
For the generation of the instance space, the features are selected through the feature selection process discussed in Section~\ref{sec:featsel} and projected to $2D$ instance space. The list of features selected in our experiments are discussed below and their correlation with the technique's performance is reported in Table~\ref{tab:feature_correlation}.

\begin{itemize}
\item \textit{Efferent Coupling (ec)} measures the external dependencies of a CUT and its high value relates to high test effort~\cite{almugrin2016using}.
\item \textit{Lines of code (loc)} measures the total lines of code in a class, excluding blank lines and comments. It is a common metric used for the assessment of the testability of a class~\cite{ferrer2013estimating}. \item \textit{Nesting depth (nd)} and \textit{number of try/catch blocks ($num\_tc$)} measure the number of blocks nested together (e.g. inside if/else, loops, try/catch etc.). 
\item \textit{Average shortest path length ($avg\_spl$)}  measures the efficiency of test data transport to the target branch and is extracted from the CFG. 
\item \textit{Mc-Cabe’s cyclomatic complexity ($cc$)} counts the number of independent paths in a CFG of a method. For class level, we use the average and standard deviation ($avg\_cc$, $std\_cc$) of this feature for all the methods in the class. 
\item \textit{Percentage of methods having $cc>10$ ($per\_cc10$)} counts the percentage of complex methods in a class that would make it harder for testing. 
\item \textit{Radius} is the minimum eccentricity of the graph, which is defined as a maximum distance of a vertex $v$ and any other vertex $u$.  We use the \textit{average radius ($avg\_rad$)} for all the CFGs generated for each method of a class. 
\end{itemize}

Once the instance space is created, we can use it to check the appropriateness of the benchmark used for the evaluation of the performance of the SBST techniques (RQ1). For this, we are interested in exploring the instance space from three perspectives: diversity, potential bias, and the presence of challenging CUTs. 

\textit{Diversity}: Ideally, the benchmark chosen for the performance assessment of testing techniques should contain sufficiently diverse structural properties to enable the strengths and weaknesses of techniques to be exposed. Without such diversity, the trustworthiness of techniques for future untested instances is necessarily limited. To anticipate the diversity of instances in the selected benchmark, we create an expanded boundary that encloses all the test instances which can empirically exist, though, might be missing from the current instance space. 

We perform boundary analysis of the generated instance space as follows: Let $\mathbb{R}^{n\times n}$ be the correlation matrix of n features. We define two vectors  $\mathbf{f}_U = \begin{bmatrix} f_{U,1} \cdots f_{U,n} \end{bmatrix}^T$ and $\mathbf{f}_L = \begin{bmatrix}f_{L,1} \cdots f_{L,n}\end{bmatrix}^T$
containing the upper and lower bounds of feature values. From these vectors, we define a vertex vector containing a combination of values from $f_U$ and $f_L$ such that only the upper or lower bound of a feature is included. For instance, $\mathbf{v}_1 = \begin{bmatrix}f_{U,1} f_{L,2} \cdots f_{L,n}\end{bmatrix}^T$ represents a vertex vector containing the maximum value of feature 1 and minimum values of all the other features. We define a matrix $\mathbf{V} = \begin{bmatrix}\mathbf{v}_1 \cdots \mathbf{v}_q\end{bmatrix} \in \mathbb{R}^{n\times q}, q = 2^n$ containing all possible vertices created by the feature combinations. The vertices in metric $\mathbf{V}$, connected by edges, define a hyper-cube that encloses all the instances in the instance space. 

Some of the vectors in $\mathbf{V}$ represent feature combinations that are unlikely to coexist. For example, if features 1 and 2 are strongly positively correlated, it is unlikely to find instances that have a high value of feature 1 and a low value of feature 2. Therefore, a vertex vector $\mathbf{v} = \begin{bmatrix}f_{U,1} f_{L,2} \cdots f_{L,n}\end{bmatrix}^T$ would be unlikely to be near any true instance. Similarly, a vertex vector cannot simultaneously contain $\{f_{U,1}, f_{U,2} \}$ or $\{f_{L,1}, f_{L,2} \}$ if feature 1 and feature 2 are strongly negatively correlated. All such unlikely vertex vectors are eliminated from $\mathbf{V}$. The edges connecting the remaining vertex vectors are then projected into $2D$ instance space using Ar from Equation~\ref{eq:optimisation}, resulting in a matrix $\mathbf{Z}$, whose convex hull now represents the mathematical boundary of the instance space. The empty regions of the instance space -- within the defined boundaries -- represent the instances that exist theoretically but missing from the current analysis, thus compromising the diversity of the used benchmark.

\textit{Bias}: The possible bias of the datasets can also be picked by carefully visualising the instance space. The regions of the instance space where the majority of the techniques perform well represent \textit{easy} instances. Reporting the performance of any newly developed technique using such instances only would not be valid evidence of its effective performance. Furthermore, evaluating a technique using a benchmark dataset containing the majority of instances having the \textit{favourable} features for that technique would represent a biased analysis. Instance space analysis is an effective framework to identify such biases in benchmark datasets.

\textit{Challenging}: A CUT is considered challenging if it is hard to cover by the majority of the portfolio techniques. A testing benchmark containing all the trivial CUTs would make all or majority of the techniques perform well, and thus unable to reveal their relative strengths or weaknesses. On the other hand, hard and challenging CUTs would unfold what makes an instance hard for testing techniques and would motivate the researchers to devise new and better techniques to address the challenging areas.

\section{Experimental Results} \label{sec:results}

This section presents experimental results based on the methodology discussed in the previous section. We present our $2D$ instance space and analyse the distribution of instances in the generated space from the perspective of diversity, bias and the challenge they pose for testing techniques. By superimposing the distribution of a technique's performance on features' distribution, we identify the features which make an instance hard/easy for a technique to test. We then examine the footprints of the portfolio techniques to gain insights into their relative strengths and weaknesses. 

\subsection{RQ1: How adequate are the commonly used benchmarks to assess the performance of SBST techniques?}
This section explores if the benchmark datasets commonly used in SBST research are adequate for the performance assessment of SBST techniques. For this study, the CUTs are extracted from two widely used benchmark datasets: SF110 Corpus~\cite{TOSEM_evaluation} and commons-collections~\cite{commons-collections}. As container classes have been widely used due to their unique properties~\cite{just2014defects4j,arcuri2008search,fraser2012seed,arcuri2010longer}, and SF110 is a state-of-the-art benchmark dataset in SBST~\cite{panichella2018large,panichella2017automated,oliveira2018mapping,fraser2014large, arcuri2017private,bruce2019dorylus,khamprapai2021performance}, the results obtained by using these two benchmarks can be generalised for commonly used benchmarks in SBST.

\begin{figure}[!t]
    \centering
    \includegraphics[width=\linewidth]{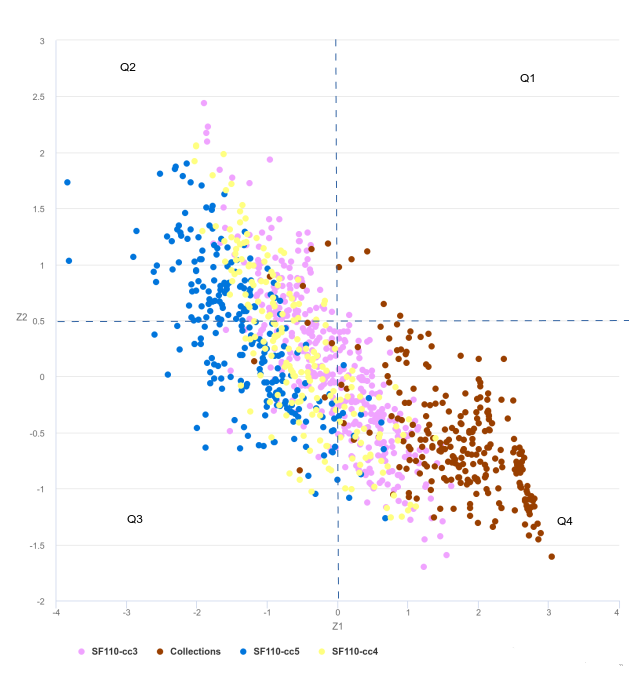}
    \caption{Instance Space for SBST techniques showing sources of CUTs. $z_1$ and $z_2$ are the coordinates in the $2D$ space as shown in Equation 1.}
    \label{fig:sources}
\end{figure}

Figure~\ref{fig:sources} shows the location of various CUTs and their sources within the instance space. The space has been divided into quadrants for ease of visualisation and reference in the later sections. SF110 is a collection of 110 open-source projects having software classes of various size and complexity. For visualisation purposes, we categorise SF110 CUTs according to their \textit{cyclomatic complexity} ($cc$). The CUTs under SF110-3 have $cc$ value in the range [3,4), SF110-4 contains CUTs in the range [4,5),  while CUTs under the category SF110-5 have $cc\geq5$. The maximum $cc$ value we observed for any CUT in our dataset is 34.5, which belongs to the project \textit{tullibee}. It should be noted that the sole purpose of categorisation on the basis of $cc$ is to show the source of instances. The instance space itself is generated based on the features selected in Section~\ref{sec:constructing_IS} and the techniques' performance based on these features.

It can be seen from Figure~\ref{fig:sources} that the instances from commons-collections are located mainly in Quadrants 1 and 4, and span only a small area of the instance space. The small area occupied by commons-collections indicates the limited diversity in the feature values of these CUTs and thus using them alone for the evaluation of SBST techniques would lack confidence in the results. However, most of these instances are not overlapped by SF110 instances, indicating their unique nature and different feature values from SF110. Therefore, it would be interesting to include these CUTs in the automated testing research along with SF110, which is the most widely used benchmark. 

On the other hand, SF110 instances cover a wide region of the instance space and are distributed across all four quadrants of the space. SF110-5 CUTs tend to cover the left half of the space, while SF110-3 and SF110-4 are dispersed across the whole space. Nevertheless, there are sparse regions of instance space having little to no instances. In particular, the instances in Quadrants 1 and 3 lack diversity as well as density. As explained in Section \ref{sec:constructing_IS}, the boundary around the instance space encloses all the possible instances calculated on the basis of minimum and maximum values of the features. The empty regions within the boundaries indicate that test instances are theoretically possible in these areas, however, are missing in the current analysis, thus compromising the diversity of the generated instance space. Albeit being outside the scope of this study, we can identify the regions of the space where the addition of new test instances would be valuable to support greater insights. New test instances with controllable properties can then be generated and added to the space to fill in the gaps~\cite{munoz2017generating}.

The visualisation of the distribution of techniques' performance in the generated instance space could give clues about the possible bias of the benchmark instances. Figure~\ref{fig:algo_coverage_dist} shows the performance distribution of portfolio techniques using branch coverage as the performance metric. It can be seen that most of the instances in the fourth quadrant are easy to cover by all the testing techniques (represented by blue colour). Even simple techniques like Random testing and Monotonic GA are effective for these instances. In our instance space, these instances mainly belong to commons-collections and SF110 having very low $cc$ values.

As discussed in Section~\ref{subsec:program-space}, the evaluation of testing techniques using container classes (e.g. commons-collections) is common as they are generic in nature, independent of programming language and free of complex environmental dependencies~\cite{andrews2011genetic, arcuri2009insight, arcuri2010longer, arcuri2008search, baresi2010testful,just2014defects4j,fraser2012seed}. However, as can be seen in the generated instance space, these instances are too easy to cover and thus are \textit{biased} toward showing good performance of almost every testing technique. Therefore, these instances alone shouldn't be relied upon for the evaluation of a testing technique. Similarly, instances from SF100 having low $cc$ values are trivial to stress test the performance of techniques and testing techniques should not be evaluated on  these CUTs alone. The idea of filtering CUTs on the basis of $cc$ and evaluating the testing technique only on non-trivial CUTs is already prevailing in the testing community. In many recent automated testing studies, the techniques have been evaluated using CUTs having $cc >= 5$ ~\cite{oliveira2018mapping, panichella2017automated, panichella2017java}.

The performance distribution across instance space also shows the regions of the space which are hard to cover by one or more techniques. The instances residing at the top of Q2 are the hardest, and none of the portfolio techniques performs well on them. On the contrary,  The instances at bottom of Q4 are easy for all the testing techniques. For the remaining instances, portfolio techniques give different performances. This gives us confidence that the benchmark used in the current study contains a combination of easy and hard instances and is capable of revealing the strength and weaknesses of the portfolio techniques. However, it can be further improved by adding more instances to fill the empty gaps and thus introducing more diversity in the space.

In summary, the answer to RQ1 is as follows:
\begin{mdframed}[leftmargin=10pt,rightmargin=10pt]

SF110, which is a widely used benchmark for SBST, contains a combination of easy and hard CUTs, that are capable of revealing the strength and weaknesses of the testing techniques. However, it lacks CUTs having diverse properties and there is potential to add more instances to this benchmark, enabling techniques to be comprehensively tested under all theoretically possible conditions. On the contrary, Java commons-collections is composed of CUTs for which all the portfolio techniques perform well, therefore it is biased towards showing the good performance of almost every testing technique. Hence, this benchmark should not be trusted alone and used in combination with other benchmarks to support the broadest possible conclusions. 
\end{mdframed}

\subsection{RQ2: What influences the effectiveness of SBST techniques?}
We analyse the distribution of the performance of portfolio techniques and identify the features of the CUTs that make it easy or hard for a testing technique, thus impacting the performance. 

\subsubsection{Distribution of the portfolio techniques in the instance space}

\begin{figure*}[!t]
    \subfloat[Random testing]{\includegraphics[width=0.32\linewidth]{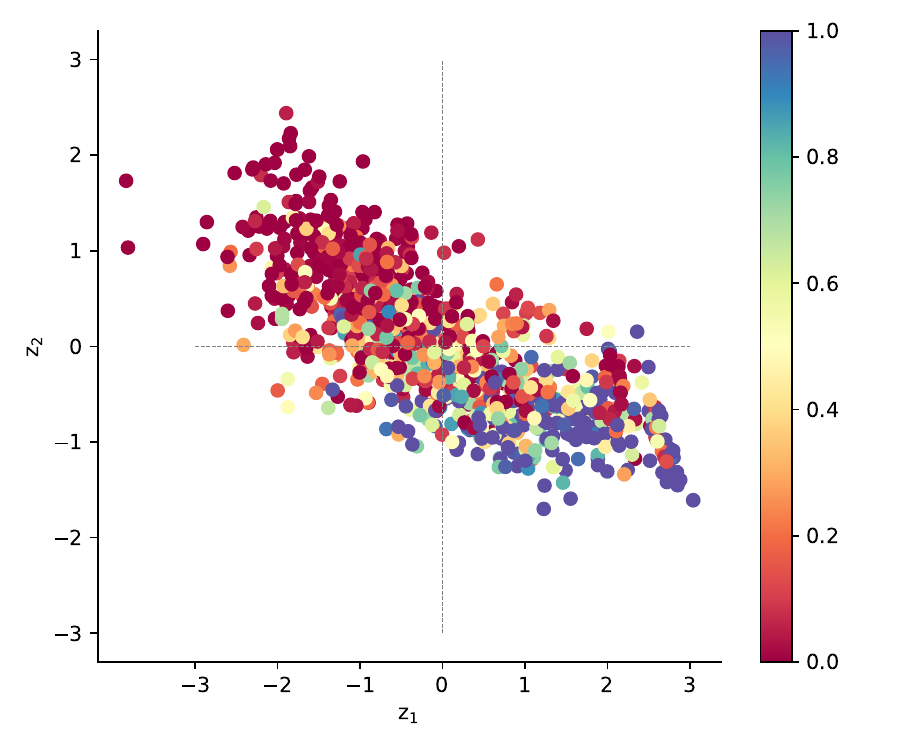}
    \label{fig:coverage_random_testing}}%
	\subfloat[Monotonic GA]{\includegraphics[width=0.32\linewidth]{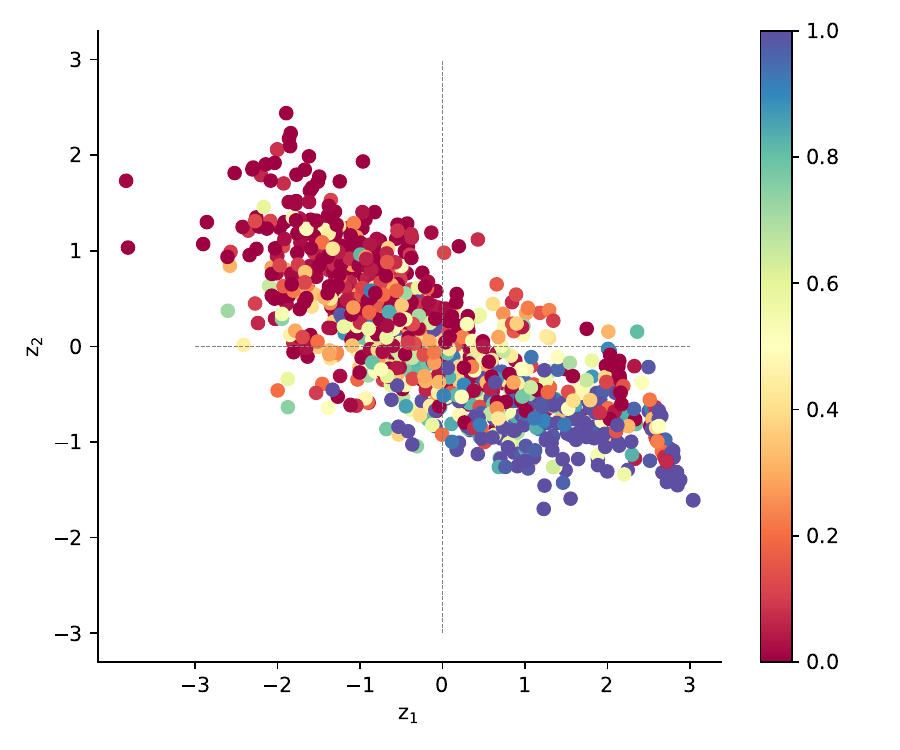}\label{fig:coverage_monotonic}}%
	\subfloat[MIO]{\includegraphics[width=0.32\linewidth]{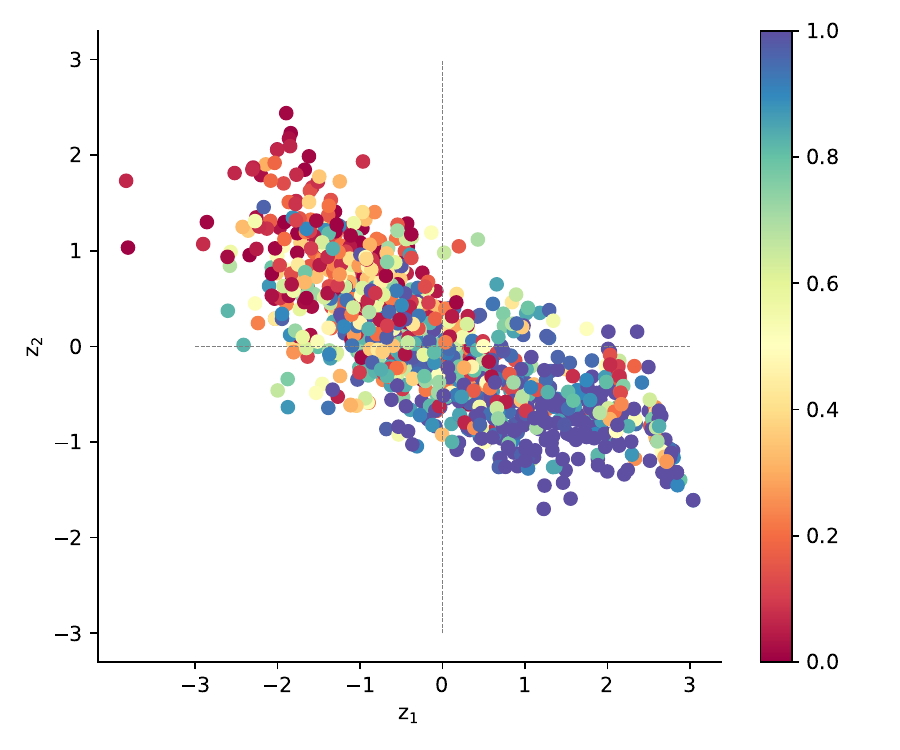}\label{fig:coverage_mio}}%
	\\
	\subfloat[MOSA]{\includegraphics[width=0.32\linewidth]{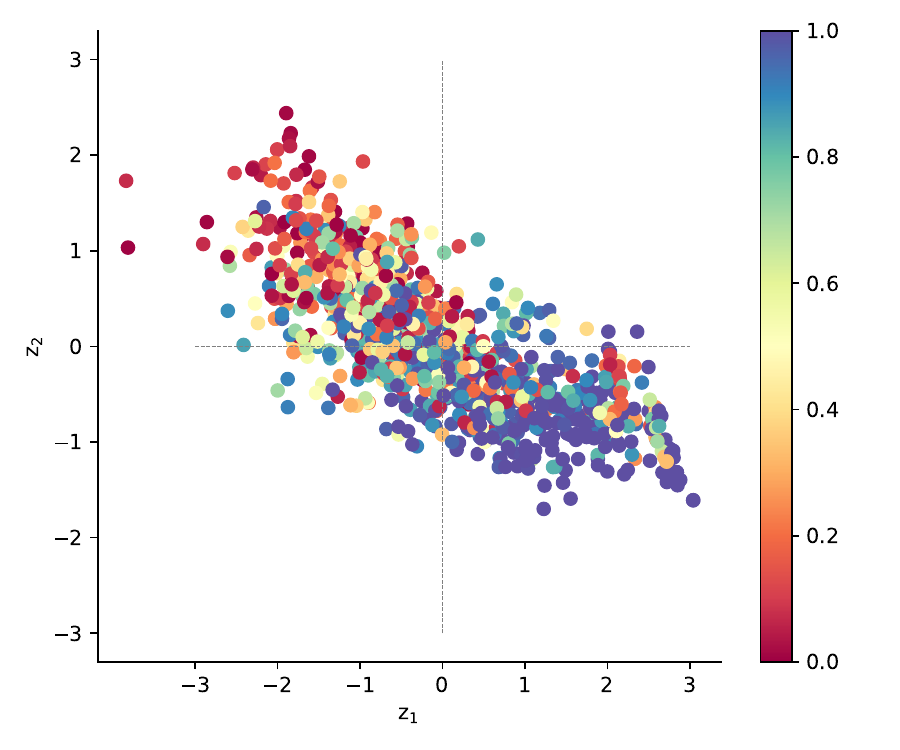}\label{fig:coverage_mosa}}%
	\subfloat[DynaMOSA]{\includegraphics[width=0.32\linewidth]{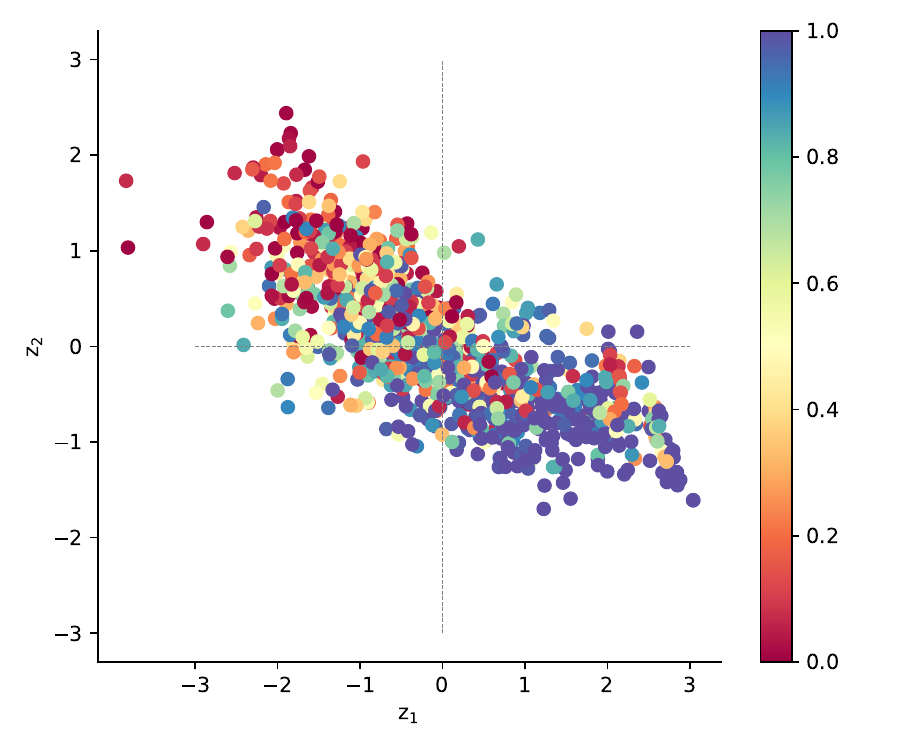}\label{fig:coverage_dynamosa}}%
	\subfloat[WSA]{\includegraphics[width=0.32\linewidth]{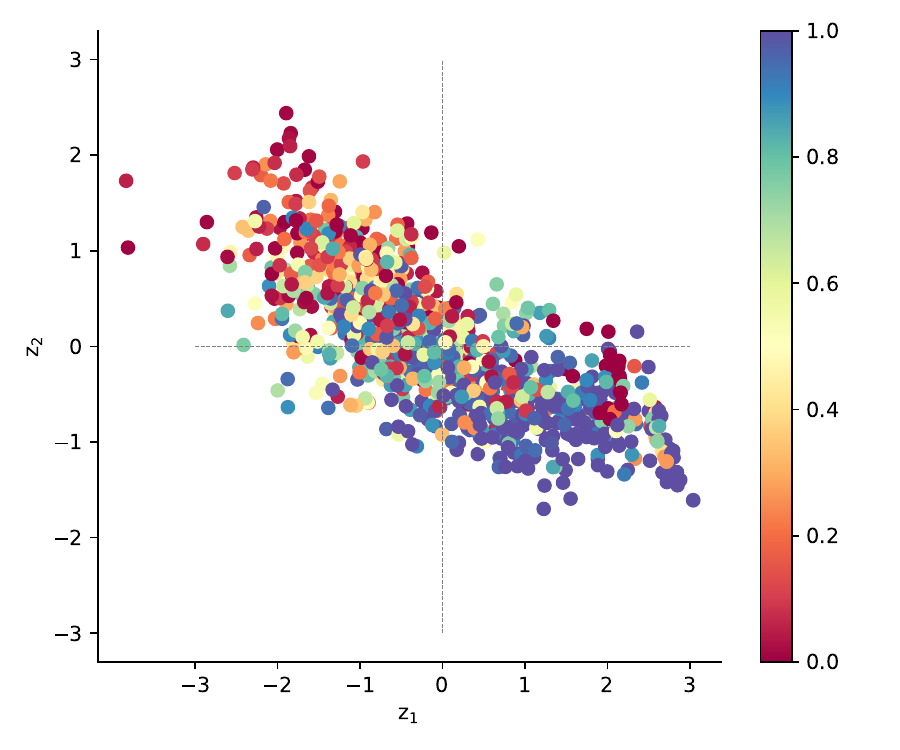}\label{fig:coverage_wsa}}%
	\caption{Distribution of the performance of SBST techniques, from minimum (red) to maximum (blue) branch coverage.}
	\label{fig:algo_coverage_dist}
\end{figure*}
Figure~\ref{fig:algo_coverage_dist} shows the distribution of the coverage achieved by the portfolio techniques on problem instances. Coverage values range between 0 and 1; shown as a colour range from red (minimum) to blue (maximum). 

It can be seen that, for most of the instances in Quadrant 1, 
MOSA, DynaMOSA, WSA and MIO achieve medium to high coverage. However, Random testing and Monotonic GA do not perform well in this area of the space. 

Quadrant 2 is occupied by the instances which are harder to cover by any portfolio technique on average. For most of the instances lying in this quadrant, the coverage achieved is very low. However, the same techniques which perform better in Quadrant 1, give comparatively higher coverage for the instances in this quadrant too. Although the \textit{number of good instances} is slightly different for each technique, the general trend of the performance is the same, i.e., MOSA, DynaMOSA, WSA and MIO perform comparatively better, while Monotonic GA and Random testing perform the worst. Furthermore, the number of good instances for MIO is lower than the other good-performing techniques. 

Quadrant 3 contains a combination of easy and hard instances, as all the techniques show medium-to-high performance for most of the instances. Random testing and Monotonic GA are exceptions, as they give reasonable coverage only on a few of the instances here. The instances where these techniques perform better lie close to Quadrant 4 (instances having low $cc$ values).

Quadrant 4 is the only part of the space where Random testing and Monotonic GA give good coverage for the majority of the instances. This quadrant consists of the instances where all 6 portfolio techniques achieve high coverage on average. However, the number of good instances is lower for Monotonic GA and Random testing compared to other techniques. If we look more carefully, these two techniques give better coverage on the part of Quadrant 4 which is occupied by the commons-collections instances. 

\subsubsection{Distribution of the selected features in the instance space}

Figure~\ref{fig:feat_dist} shows the distribution of the selected features across the instance space. The feature values show the following trends:

\begin{figure*}[!t]
    \subfloat[avg\_spl]{\includegraphics[width=0.32\linewidth]{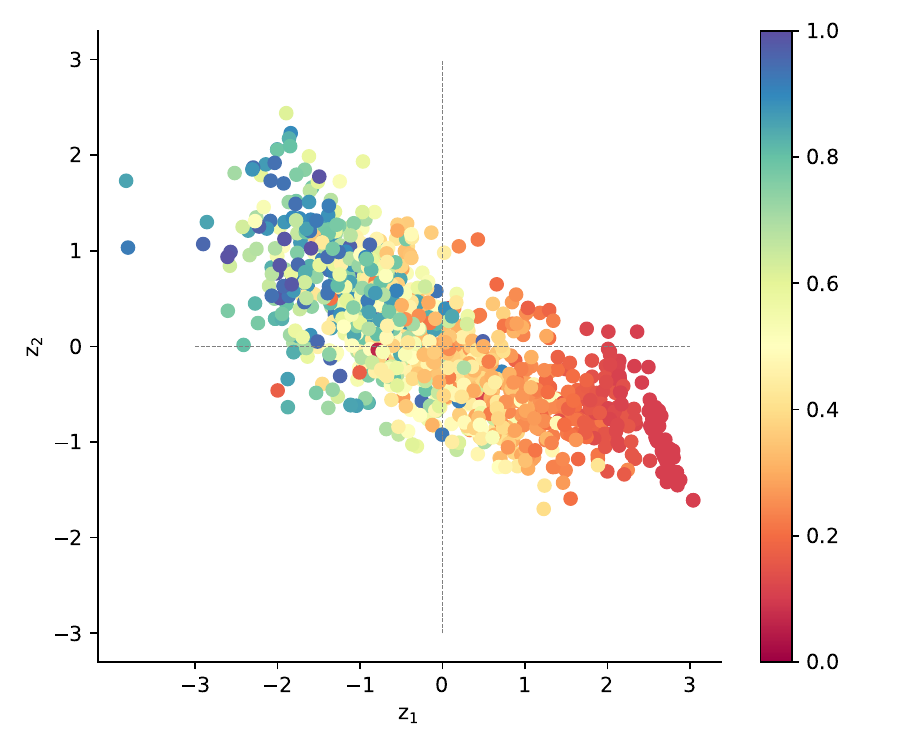}\label{fig:avg_spl}}%
	\subfloat[std\_cc]{\includegraphics[width=0.32\linewidth]{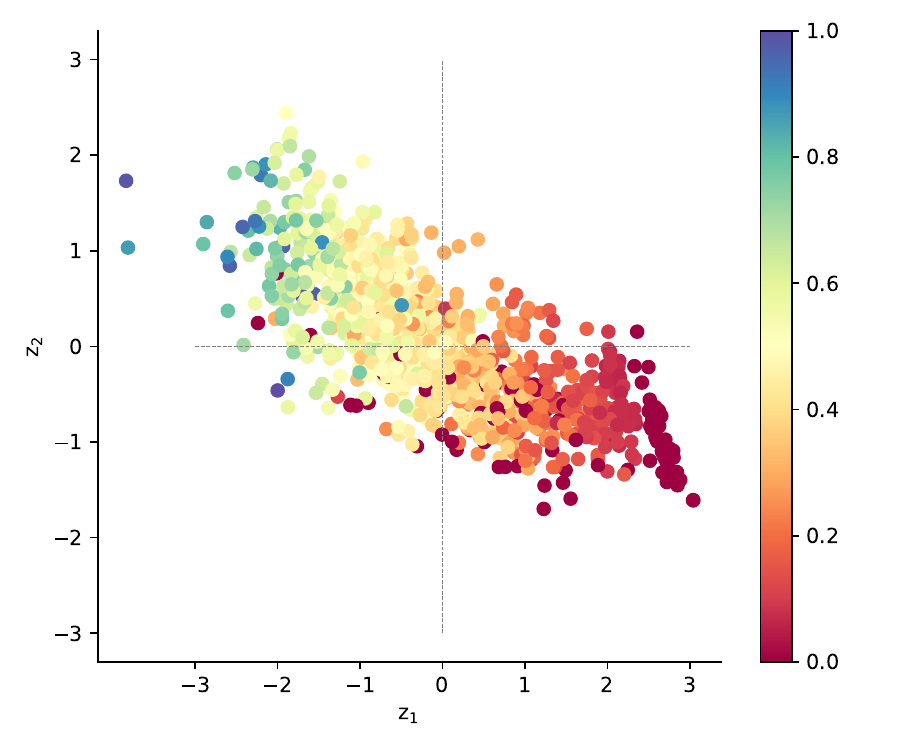}\label{fig:std_cc}}%
	\subfloat[loc]{\includegraphics[width=0.32\linewidth]{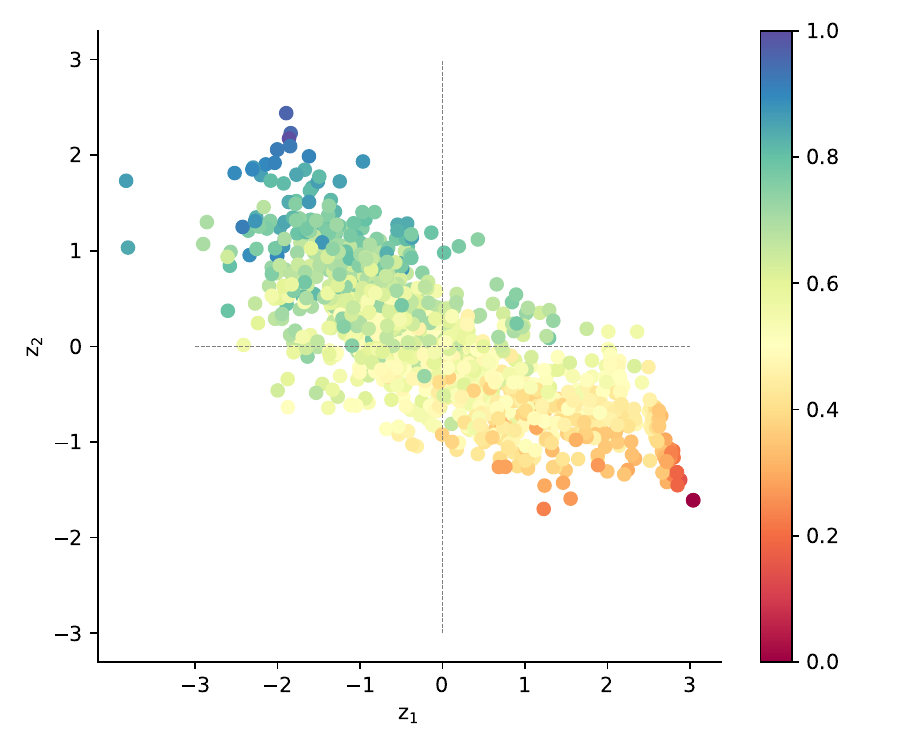}\label{fig:loc}}\\%
	\subfloat[nd]{\includegraphics[width=0.32\linewidth]{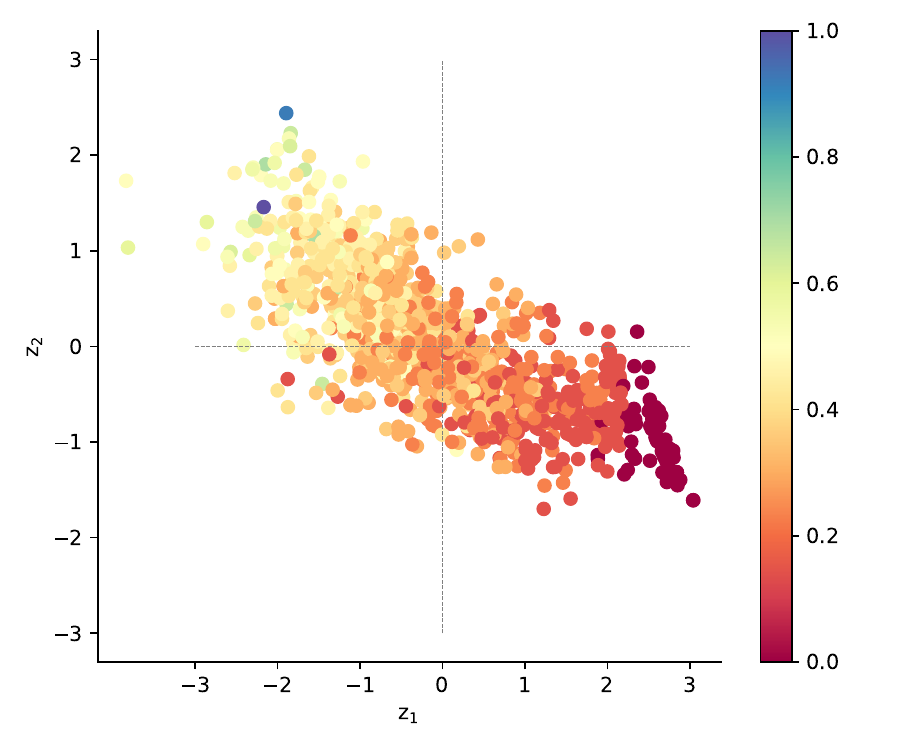}\label{fig:nd}}%
	\subfloat[num\_tc]{\includegraphics[width=0.32\linewidth]{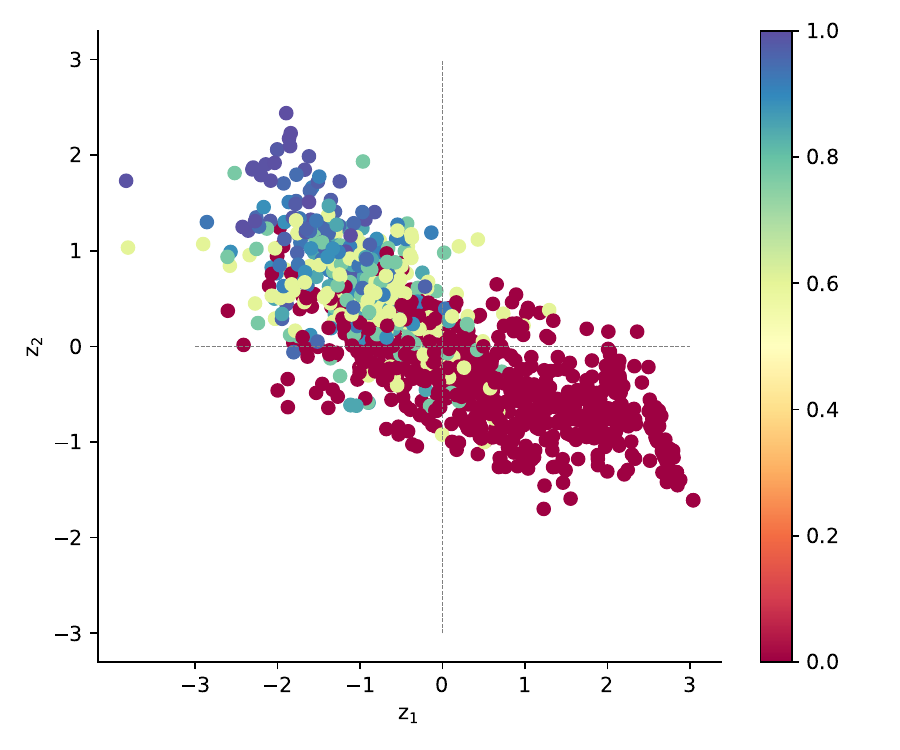}\label{fig:num_tc}}%
	\subfloat[ec]{\includegraphics[width=0.32\linewidth]{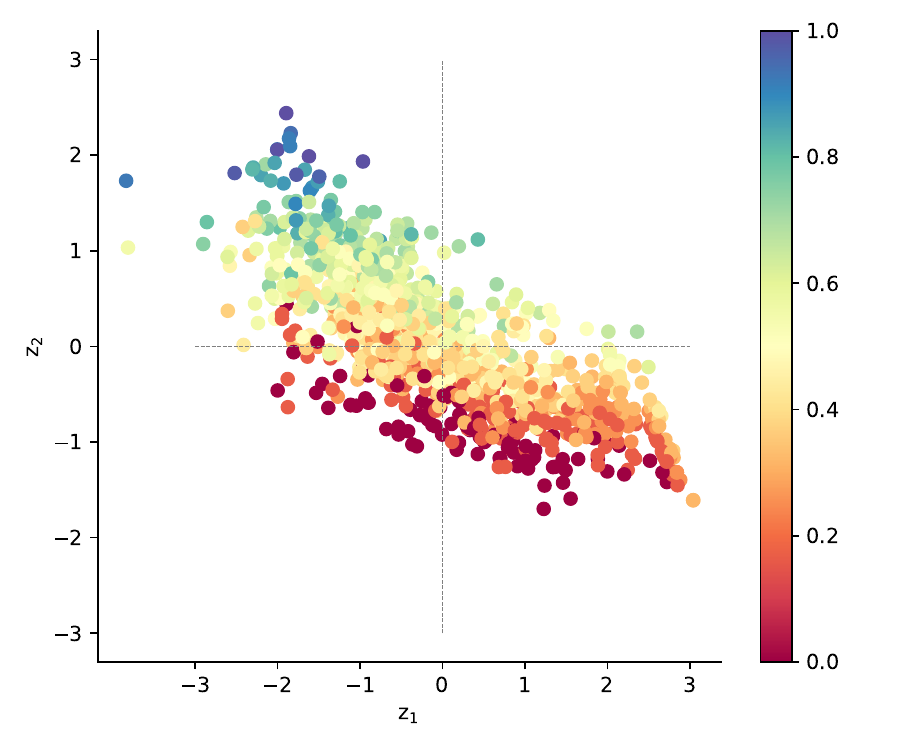}\label{fig:ec}}\\%
	\subfloat[per\_cc10]{\includegraphics[width=0.32\linewidth]{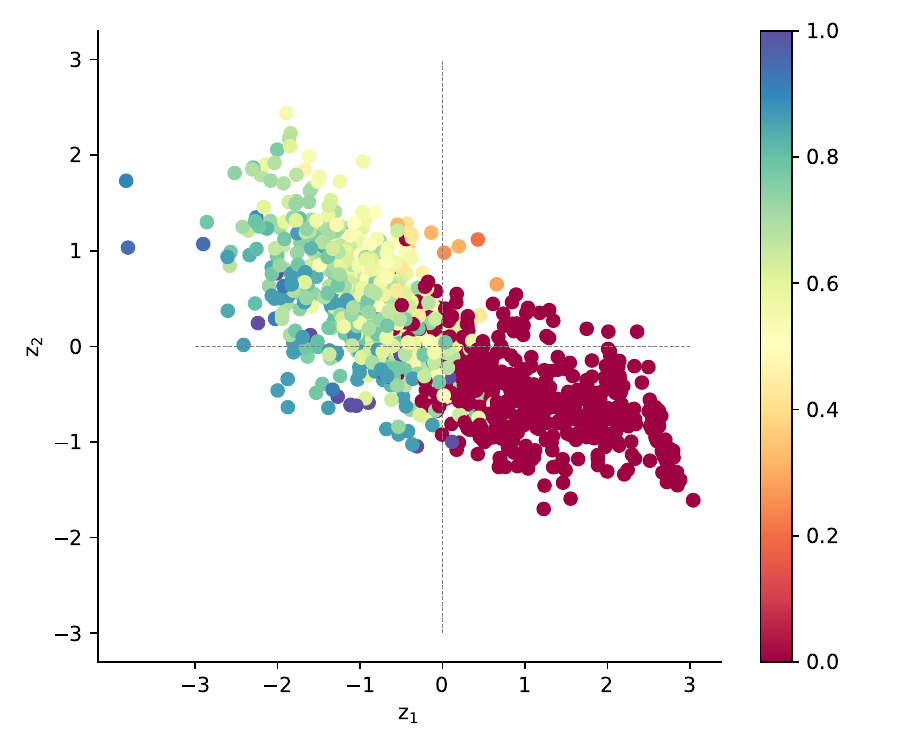}\label{fig:per_cc10}}%
	\subfloat[avg\_cc]{\includegraphics[width=0.32\linewidth]{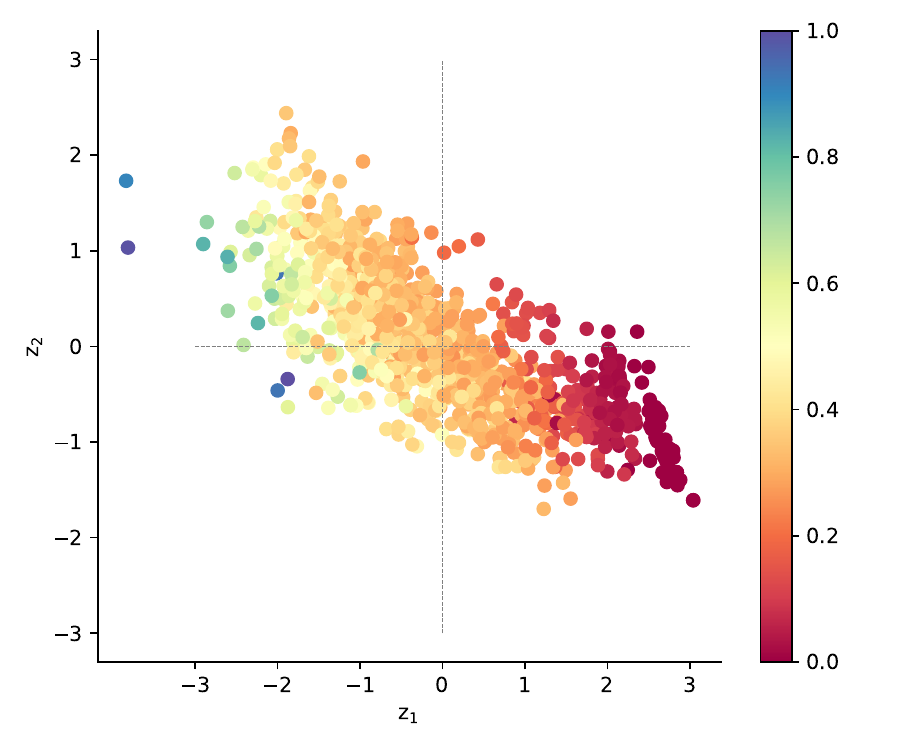}\label{fig:avg_cc}}
	\subfloat[avg\_rad]{\includegraphics[width=0.32\linewidth]{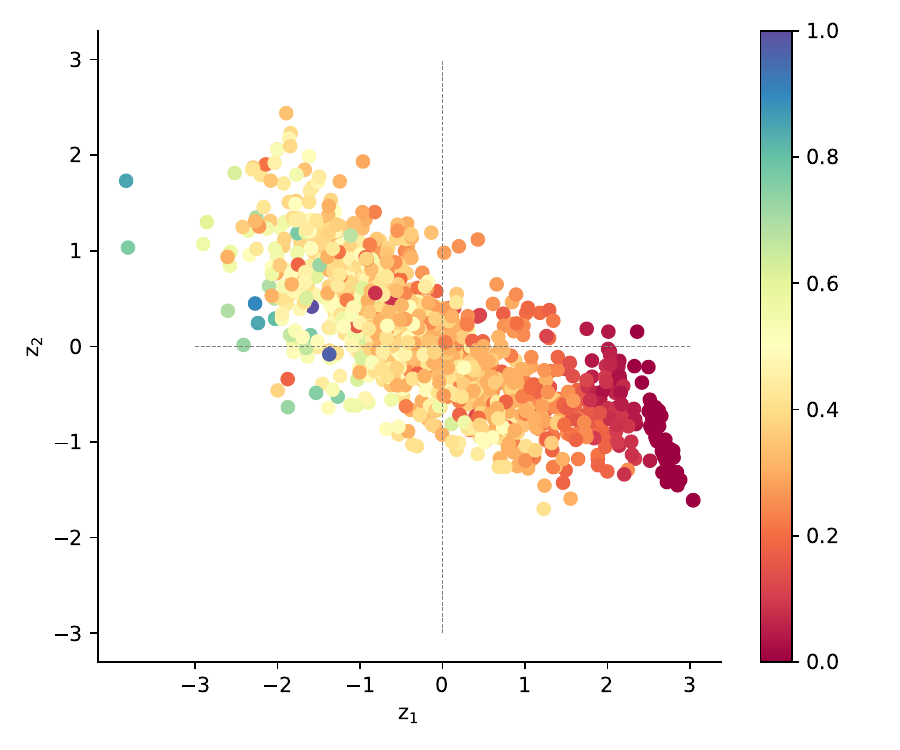}\label{fig:avg_rad}}%
	\caption{Distribution of meta-features, from minimum (red) to maximum (blue) values.}
    \label{fig:feat_dist}
\end{figure*}

\begin{itemize}
    \item \textit{Average shortest path length (\textit{avg_spl}):} decreases from top to bottom (diagonally), showing that commons-collections have the shortest observed path while the instances from SF110-5 located in Quadrant 2 have the largest observed path.  
    \item \textit{Standard deviation Mc-Cabe’s cyclomatic complexity (\textit{std $cc$}):} shows a decreasing trend from top to bottom. Instances in the first and fourth quadrants show very low to low values (the lowest values are at the bottom of Quadrant 4 which are instances from commons-collections) while these values are medium to high for the instances in Quadrants 2 and 3.
    \item \textit{Lines of code (\textit{loc}):} has a similar trend as \textit{std $cc$}. However, the number of instances having a high value of this feature is quite large, while the number of instances having a very low value is small comparatively.
    \item \textit{Nesting depth (\textit{nd}):} most of the instances have very low to low values for this feature. It can be seen that commons-collections have the lowest nesting depth, while most of the SF110 instances have low to medium values. It can be noted that the instances which are deeply nested are rare. Deep nesting makes an instance harder to test as the code elements deeper in the nesting structure are harder to reach, depending on how easy/hard the outer targets are to cover. Therefore, an addition of such instances in the benchmarks would challenge the performance of testing techniques~\cite{mcminn2009empirical}. 
    \item \textit{Number of try/catch blocks (\textit{num_tc}):} decreases from top to bottom, and divides the instance space more or less horizontally. 
    \item \textit{Efferent coupling (\textit{ec}):} This feature shows a decreasing trend from top to bottom of the instance space diagonally. The instances in Quadrant 2 have the highest value for this feature, while first and fourth quadrant got instances having medium to high efferent coupling. This shows an interesting fact that although the instances from commons-collections (located mainly in Quadrant 4) are easy in terms of \textit{avg_$cc$}, \textit{nd} etc., their \textit{ec} values are still medium to high. This depicts the fact that these classes are tightly bound to other classes, the phenomenon that can make these apparently easy instances (due to low values of above-mentioned features) hard to cover for some techniques.
    \item \textit{Percentage of methods having $cc > 10$ (\textit{per_$cc10$}):} The instances in the second and third quadrants show medium to high values, while instances in the right half contain the majority of methods having low $cc$ values.
    \item \textit{Average Mc-Cabe’s cyclomatic complexity (\textit{avg_$cc$}):} commons-collections and other instances in the first and fourth quadrants have low \textit{avg_$cc$}, while Quadrants 2 and 3 contain instances having medium to high \textit{avg_$cc$} values. 
    \item \textit{Average radius (avg_rad): } Most of the instances at right side of the instance space have low value for this feature, while instances in Quadrant 2 and 3 shows medium to high values. 
\end{itemize}

By superimposing Figure~\ref{fig:algo_coverage_dist} on Figure~\ref{fig:feat_dist}, we can map the performance of the techniques to the features of the instances. The instances in Quadrants 1 and 4 are easy from a testing perspective as they represent simple CUTs having lower values for  \textit{num_tc}, \textit{nd}, \textit{std\_cc} and \textit{per_cc10}. However, Quadrant 1 is also characterised by medium to high values of efferent coupling that make these CUTs dependent on other CUTs for their effective testing. On the contrary, the efferent coupling is low for the instances in Quadrant 4.

By scanning the distribution of the above-mentioned features and techniques' performance, we can infer that Random testing and Monotonic GA perform as good as sophisticated multi-objective and multi-population search techniques for the simple instances defined in terms of the features mentioned above. Most of these instances belong to commons-collections. Thus, our instance space confirms the findings of Shamshiri et al.~\cite{shamshiri2015random} that, for the unit test generation, Random testing is as good as sophisticated search-based techniques. It is important to mention here that the testing benchmark used in the study by Shamshiri et al.~\cite{shamshiri2015random} consists of container classes from C programming language, while commons collections contain similar classes written in Java. As discussed before, container classes are the implementation of data structures like lists, arrays etc. and thus their implementation is quite similar across different languages. 

However, even for commons-collections instances, the performance of these two techniques is badly impacted by higher values of efferent coupling. Thus, software classes that depend on other classes for their proper operation are harder to cover by these two techniques. Although to a lesser extent, efferent coupling is a feature that negatively impacts the performance of all the testing techniques. 

All the techniques in quadrant 2 fail to achieve good coverage on most of the instances located here. As shown in Figure~\ref{fig:feat_dist}, these instances are characterised by high values of \textit{ec}, \textit{num_tc}, \textit{avg_spl} and medium to large \textit{loc}, which are the most prominent features which explain what makes a CUT harder to get covered by search-based software testing techniques. A closer look at the performance in this quadrant shows that the coverage achieved by the techniques is medium to high for a small set of instances that are located close to the origin. Even Random testing and Monotonic GA achieve a reasonable coverage for some instances in this area (the presence of blue, green and yellow points). These instances are characterised by medium values of \textit{loc} and very low \textit{num\_tc}. Therefore, even with medium to high \textit{ec}, smaller CUTs in terms of size (\textit{loc}) gain higher coverage. A possible explanation for this observation is that the smaller CUTs have lower number of testing goals and thus larger search budget is available for each target, giving enough time to the testing techniques to cover even the harder targets (in terms of \textit{ec} or \textit{avg\_cc} etc.). MOSA, DynaMOSA, and WSA produce the best performance in this area, followed by MIO that shows good performance for comparatively lesser number of instances. 

The instances in Quadrant 3 are characterised by many methods having $cc > 10$. However, this features doesn't seem to impact the relative performance of the techniques in this quadrant. MOSA, DynaMOSA, and WSA give medium to high coverage, while Random testing and Monotonic Ga produce mixed coverage (a combination of low, medium and high values). It is very hard to identify the defining features of the instances in this quadrant and to map them to the performance of the techniques. This region of the generated instance space, therefore, doesn't provide any interesting information about the impact of CUTs features on a technique's performance.

From all the observations reported in this section we infer that:
\begin{mdframed}
The most salient features which define the degree of hardness of a CUT from a testing perspective are \textit{ec}, \textit{num_tc}, \textit{loc} and \textit{avg_spl}. Among these features, \textit{ec} is the most prominent feature that impacts the performance of the investigated SBST techniques. Even a simple CUT, characterised by small size and fewer number of try/catch blocks, becomes hard to cover if it is tightly coupled to other classes. 
\end{mdframed}

\subsection{RQ3: What are the strengths and weaknesses of existing SBST techniques?}

\begin{table*}[!t]
	\centering
	\caption{Footprint analysis for the portfolio techniques, including their area $\left(\alpha_{N}\right)$, density $\left(d_{N}\right)$ and purity $\left(p_{N}\right)$, whenever the technique is good or best, identified with the subscripts $G$ or $B$ respectively.}
	\begin{tabular}{lrrrrrr}
		\toprule
		             & $\alpha_{N,G} (\%)$ & $d_{N,G}\%$ & $p_{N,G}\%$ & $\alpha_{N,B}\%$ & $d_{N,B}\%$ & $p_{N,B}\%$ \\ \midrule
		MOSA         &                63.7 &       105.7 &        83.9 &             24.9 &        85.1 &        86.3 \\
		DynaMOSA     &                48.1 &       105.3 &        85.0 &              1.6 &        87.8 &          80 \\
		WSA          &                43.2 &       129.1 &        80.9 &              1.6 &        87.9 &          80 \\
		MIO          &                33.5 &        89.2 &        84.3 &              0.3 &        83.2 &         100 \\
		Monotonic GA &                 3.8 &        90.6 &        97.3 &              0.0 &         0.0 &         0.0 \\
		RT           &                 1.3 &       128.9 &        94.4 &              0.0 &         0.0 &         0.0 \\ \midrule
		Average      &                32.2 &       108.1 &        87.6 &              4.7 &        57.3 &        57.7 \\ \bottomrule
	\end{tabular}%
	\label{tab:footprint_analysis}%
\end{table*}
\begin{figure*}[!t]
    \flushleft
    \subfloat[][Random testing]{\includegraphics[width=0.32\linewidth]{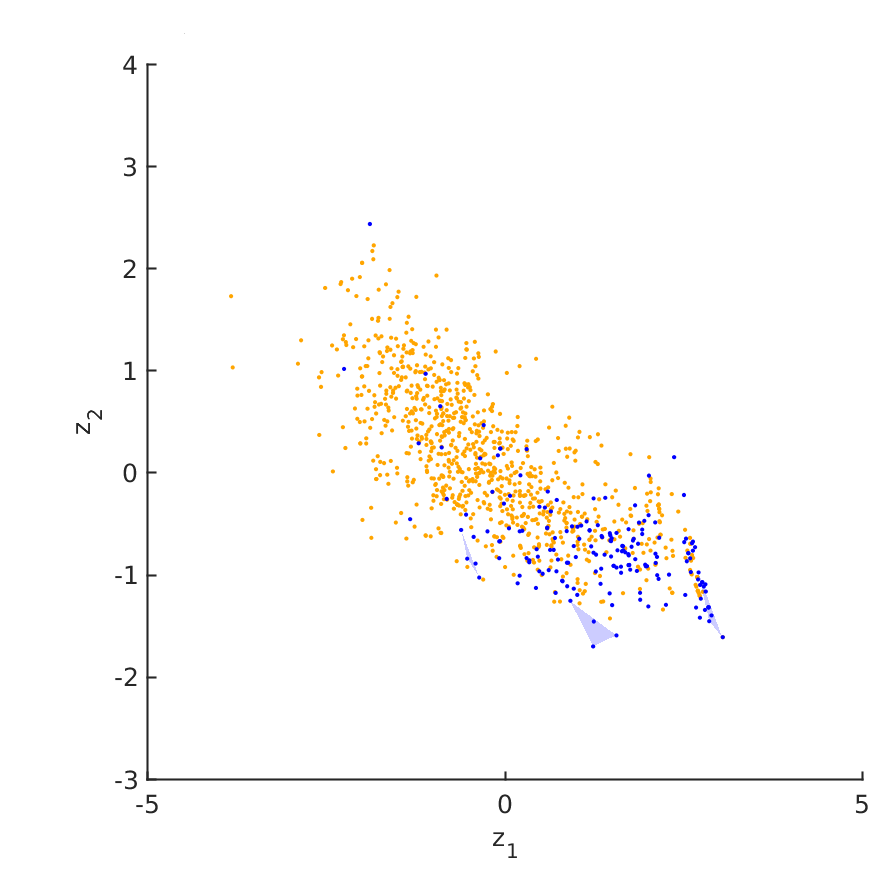}\label{fig:footprint_random_testing}}%
	\subfloat[][Monotonic GA]{\includegraphics[width=0.32\linewidth]{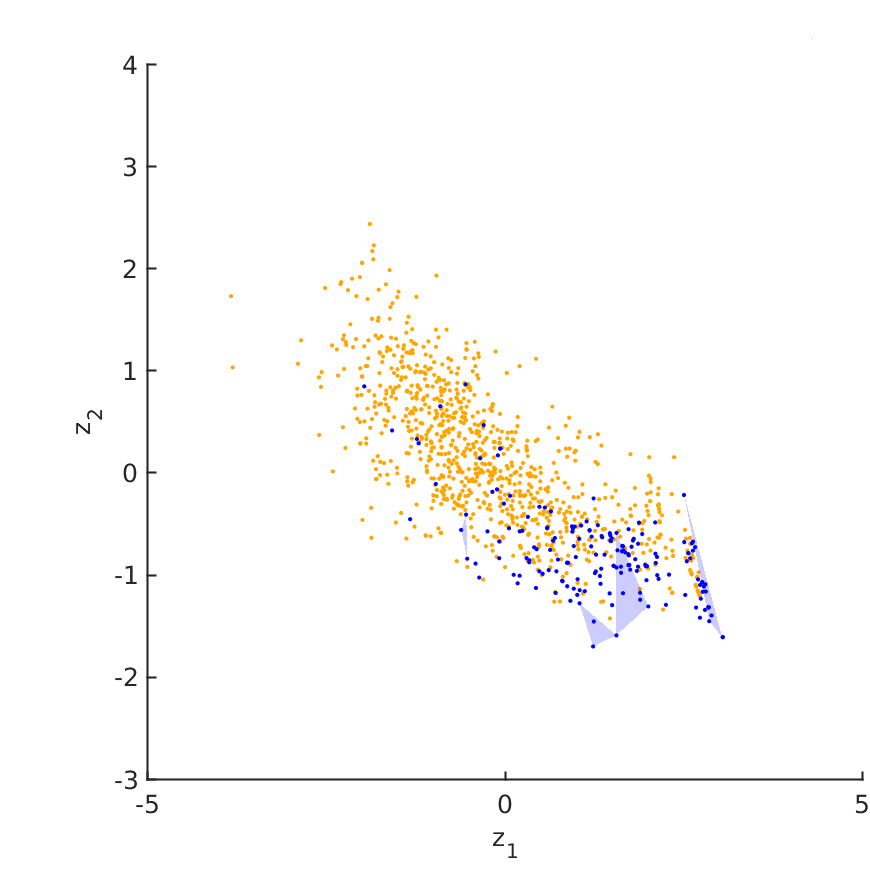}\label{fig:footprint_monotonic}}%
	\subfloat[][MIO]{\includegraphics[width=0.32\linewidth]{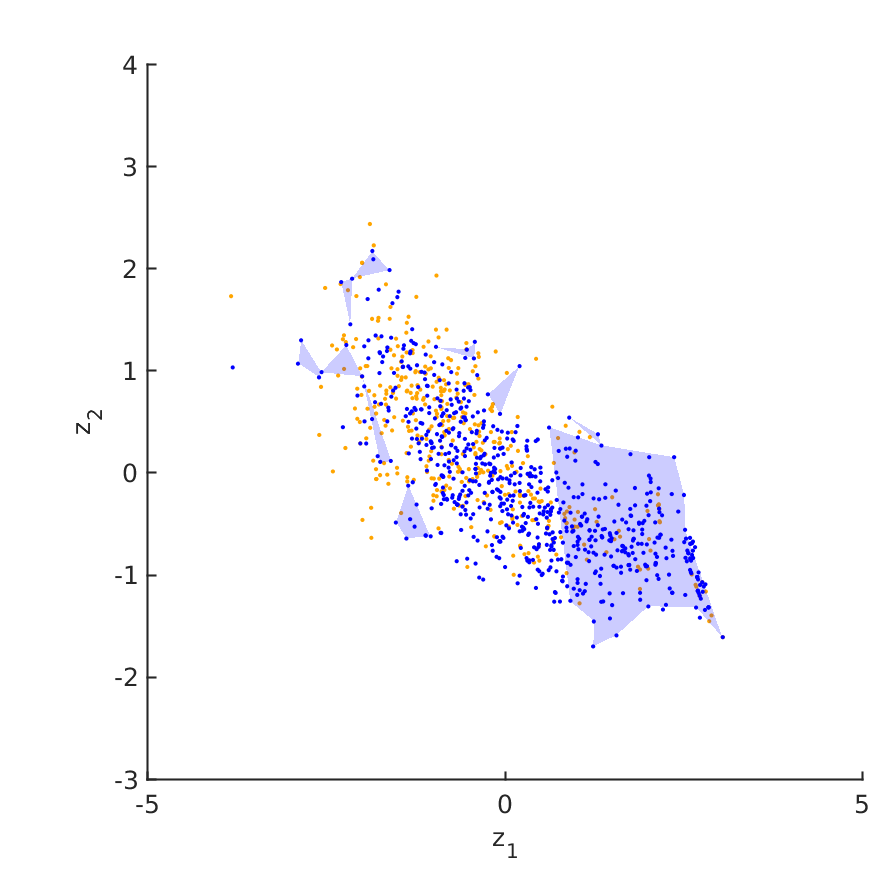}\label{fig:footprint_mio}}%
	\\
	\subfloat[][MOSA]{\includegraphics[width=0.32\linewidth]{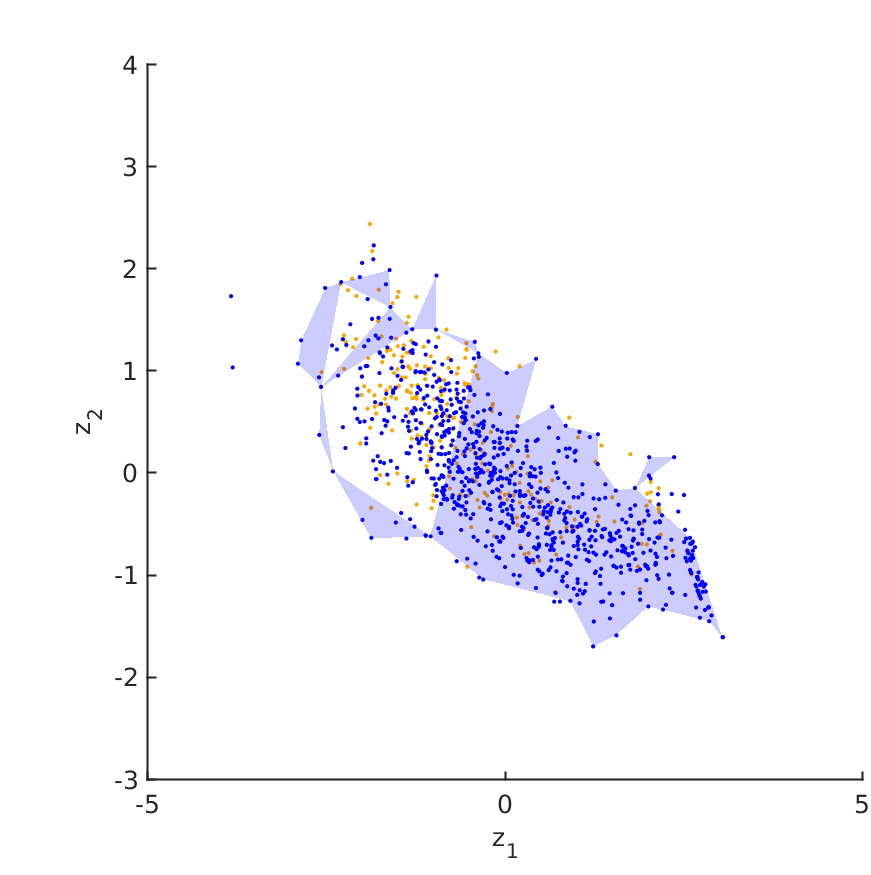}\label{fig:footprint_mosa}}%
	\subfloat[][DynaMOSA]{\includegraphics[width=0.32\linewidth]{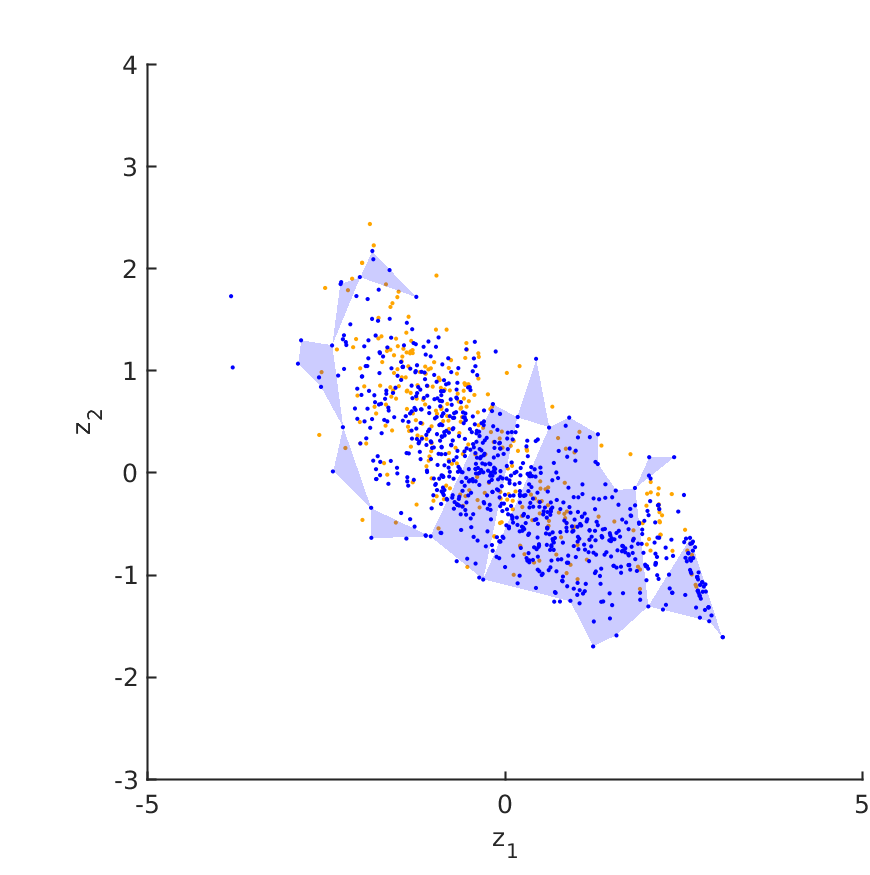}\label{fig:footprint_dynamosa}}%
	\subfloat[][WSA]{\includegraphics[width=0.32\linewidth]{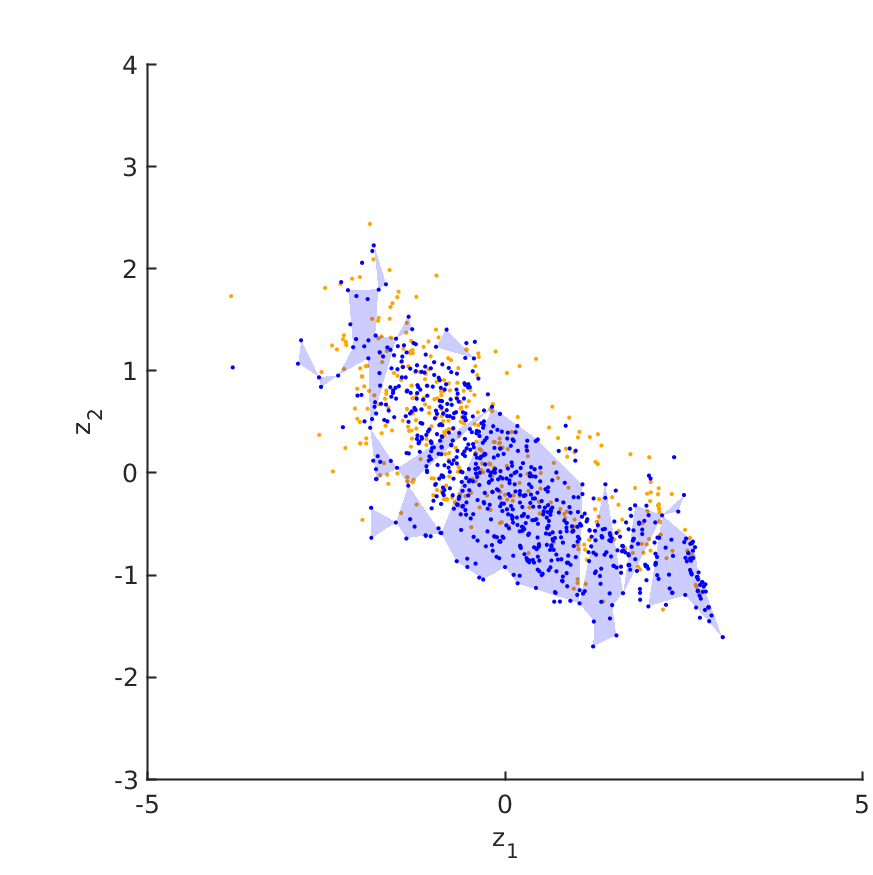}\label{fig:footprint_wsa}}%
	\caption{Footprints (regions with blue background) for (a) Random testing, (b) Monotonic GA, (c) MIO, (d) MOSA, (e) DynaMOSA, and (f) WSA, where good performance represents the coverage achieved by a technique within 5\% of the best coverage value on an instance. }
	\label{fig:algo_footprints}
\end{figure*}

Table~\ref{tab:footprint_analysis} reports the footprints of the portfolio techniques defined by area ($\alpha_{N}$), density ($d_{N}$) and purity ($p_{N}$), along with the subscripts G or B showing that the technique is good (as per the definition in Section~\ref{sec:configuration}) or best (among all the portfolio techniques). These values are normalised over the total density ($d_s$) and total area ($\alpha_s$) of the convex hull containing all instances. The individual footprints of the portfolio techniques are shown in Figure~\ref{fig:algo_footprints}. 

Although MOSA, DynaMOSA and WSA show similar performance in terms of average coverage (Fig.~\ref{fig:algo_coverage_dist}), their footprints reveal their differences. MOSA covers the largest area with the second-lowest purity 
$\left(\alpha_{N,G}=63.7\%,p_{N,G}=83.9\%\right)$. DynaMOSA on the other hand, occupies a smaller region of the space ($\alpha_{N,G}=48.1\%$), however, with slightly higher purity ($p_{N,G}=85.0$). It is not surprising that MOSA and DynaMOSA share many regions of the space under their footprints, as the latter is an extension of the former that addresses the problem of dependence of the testing targets on their parent branches (for instance, in deeply nested loops). As there are very few instances in our space having deeply nested branches (Fig.~\ref{fig:nd}), the performance of DynaMOSA would be very similar to MOSA. However, the region in Quadrant 4 where instances from commons-collections have higher \textit{ec}, illustrates a difference between these two techniques, as it is covered by MOSA but not by DynaMOSA's footprint. Therefore, MOSA is more effective in covering instances with higher \textit{ec} compared to DynaMOSA. Furthermore, The area of the space where MOSA performs the best ($\alpha_{N,B}$) is also much higher compared to all other techniques. WSA occupies the smallest area (43.2\%) among MOSA, DynaMOSA, and WSA, however, the density of its footprint is higher than average (129.1\%). Both area and density for MIO is lesser than MOSA, DynaMOSA, and WSA $\left(\alpha_{N,G}=33.5\%,d_{N,G}=89.2\%\right)$.

Random testing and Monotonic GA cover the smallest areas of the instance space (1.3\% and 3.8\% respectively), however, with high purity. The reason behind their high purity is the area of these footprints that is very small and occupies only a few instances which are easy to be covered by these techniques. Thus, the good performance of these techniques is limited to very small but pure areas in Quadrant 4. From the above footprint analysis, we infer that:

\begin{mdframed}
Multi-objective and multi-population search-based techniques are more effective than Random testing and single-objective search. Among good performers, MOSA is the most effective testing technique, as its footprint occupies the largest area of the space and is reasonably dense. MOSA's footprint also covers the area of Quadrant 4 having high efferent coupling, where other portfolio techniques could not perform very well.
\end{mdframed}

These results are in line with the findings of comparative studies, where MOSA and DynaMOSA are reported to perform better than other portfolio techniques using ''average coverage" as a performance metric~\cite{panichella2015reformulating,panichella2017automated,panichella2018large}. However, footprint analysis provides a deeper understanding of what makes an instance easy/hard for a technique compared to others, and thus \textit{explains} the relative performance of the testing techniques. 

\section{Threats to Validity}\label{sec:threats}
\textbf{Threats to internal validity} concern factors that influence the presented results. One such threat comes from the choice of features. The choice of features has a great impact on the final footprint of techniques. Furthermore, the selection of features has a significant impact on the general characteristics of the instance space, e.g. diversity, bias and boundary analysis. Therefore, feature selection is at the heart of ISA and must be chosen carefully.

We have mitigated this threat by selecting an extensive set of features extracted from the control flow graph of the CUT, software quality metrics and CUT's static features. These features are widely and effectively used in automated testing research for the assessment of testability and other purposes.

Similarly, the choice of the dataset, portfolio techniques and performance metric have a critical influence on the final results. For instance, if we change the correlation threshold to filter the features in the preprocessing stage, or change the definition of good, different features might get selected, and thus the axes of the instance space system might change, resulting in different results. However, comparing different definitions or configuration parameters is not our focus in this research and interested user can easily reproduce our experiments and play around with the configuration parameters using the code and meta-data available online \footnote{\url{https://matilda.unimelb.edu.au/matilda/problems/sbse/ast}}. Furthermore, we use the Instance Space Analysis to gain visual insights when considering the settings adapted in Section~\ref{sec:exp_design}, knowing that the methodology is general, scalable and repeatable, and final results can vary by selecting different features, techniques, dataset and performance metric.

\textbf{Threats to external validity} affect the generalisation of the results. We used a benchmark of 1088 classes chosen from SF110 and commons collections. Both of these datasets have been widely used in previous work on automated testing. Furthermore, we also made sure that we use a combination of easy and hard instances to make the instance space more diverse. 

\textbf{Threats to construct Validity} concern the relation between experimentation and theory. We have compared the performance of portfolio techniques based on \textit{branch coverage}, which is a widely used performance metric in the literature. However, the results could be more interesting by evaluating the performance based on other metrics, e.g. mutation score, and comparing the footprints of the techniques based on two or more metrics. We are very interested in such research as future work. 

\section{Related Work}\label{sec:related-work}
Automated testing techniques chosen as portfolio techniques in our proposed methodology are discussed in detail in Section \ref{sec:algospace}. Below we explore the studies that evaluate the performance of these techniques and report how the current study is different from them.

Fitness landscape analysis is performed by analyzing the distribution of fitness values of candidate solutions in the search space~\cite{varshney2013search}. The height of each point represents the fitness value; the highest point represents the best solution to the problem. The visualisation of the fitness landscape and understanding the relationship between landscape features and the performance of optimisation methods can provide valuable insights as to which method will work best for the given problem. Examples of landscape features are the size, number and distribution of the optima, the location of the global optimum and plateaus. The presence of a single optimum makes a problem easy to solve by the deterministic hill climbing method. On the contrary, the presence of multiple plateaus and local optima make it harder to be solved by a search algorithm~\cite{aleti2017analysing}. There are many studies that perform fitness landscape analysis for search-based software testing~\cite{varshney2013search,aleti2017analysing,albunian2020causes,shamshiri2015random,ewallroles}. Unlike fitness landscape analysis that defines the search problem using landscape features, our proposed methodology finds the relationship between software features and the performance of the test generation techniques. Although some of these features can be translated to landscape features, for instance, boolean branches, number of gradients etc., most of the features are generic and represent the problem instances as \textit{software classes}. Furthermore, the proposed technique can also be applied to testing methods other than search-based testing.

The performance comparison of automated testing techniques has been the focus of many studies. Ferrer et al. evaluated the performance of four multi-objective optimisation techniques, MOCell, NSGA-II, SPEA2, and PAES; two mono-objective techniques GA and ES; and two random techniques on the basis of three quality indicators: the HV, the 50\% EASs, and the maximum coverage~\cite{ferrer2012evolutionary}. In ~\cite{fraser2012whole}, Rojas et al. report the superiority of evolving a whole test suite (WS) over using a single test as an individual in the search process, while the performance of the achieved version of WS is evaluated in ~\cite{rojas2017detailed}. Panichella et al. propose a multi-objective search strategy called MOSA (Many-Objective Sorting Algorithm) and compare its performance with WS on the basis of branch coverage and search convergence time~\cite{panichella2015reformulating}. A variant of MOSA called DynaMOSA is proposed by Panichella et al. ~\cite{panichella2017automated} and the performance of the proposed technique is compared with MOSA, WS and WSA on the basis of the branch, statement and strong mutation coverage. In order to address the scalability issue of multi-objective search techniques when the number of targets (branch, statement etc) are very high, Arcuri et al. present a technique called Many Independent Objects (MIO) and reported its superior performance as compared to MOSA, Random testing and WS based online and statement coverage~\cite{arcuri2017many}. Another detailed comparative study is presented by Panichella et al. comparing the performance of various multi-objective and single-objective testing techniques~\cite{panichella2018large}. The study reported that multi-objective techniques are more effective as compared to single-objective, particularly for complex classes. Furthermore, among multi-objective techniques, DynaMOSA performs significantly better than its predecessors (MOSA and WS). The working and search strategies of all the techniques mentioned in the above paragraphs are discussed in detail in Section \ref{sec:algospace}.

Comparative studies are also available for automated test-generation tools. Cseppento et al. evaluated various Symbolic Execution-based Test Tools ~\cite{cseppento2015evaluating}. Wang et al. compared Crasher, TestGen4j, and JUB based on their mutation scores~\cite{wang2009comparison}. Chitirala et al. compared the performance of EvoSuite and Tpalus using code coverage, mutation score and size of the test suite as performance metric~\cite{chitirala2015comparing}. P{\"a}ivi and M{\"a}ntyl{\"a} et al. proposed a 12-step process to choose the right test automation tool ~\cite{raulamo2017choosing}. In~\cite{ramler2013replicated}, manual and tool-assisted testing is compared and it is reported that automated unit testing is as effective as manual testing under severe time conditions. Fraser et al. evaluate the performance of EvoSuite based on average branch coverage~\cite{fraser2014large}. 

All the studies mentioned above evaluate testing techniques/tools based on their average performance (average coverage, average mutation score etc.). Such averages don't give much information about what makes a particular technique perform better on the used dataset. Unlike these studies, our proposed methodology maps the performance of techniques to the unique features of the software and explores how these features impact the performance of one technique versus others.

There is also rich literature on accessing the testability and quality of software using software metrics. Lammermann et al. explore the relationship between software quality measures and evolutionary testing of procedural code ~\cite{lammermann2008evaluating}. As the first step, software artefacts are sorted in order of the code coverage achieved on them. These are then sorted in order of values of selected software metrics (lines of code, cyclomatic complexity etc.). The difference in the sorting order of code coverage and software metrics is used to estimate how effective is a particular metric in defining the testability of software. In ~\cite{da2017empirical}, the correlation between CK metrics~\cite{chidamber1994metrics} and line, branch and mutation coverage is reported using Spearman’s rank order. The impact of software testability metrics
at various stages of the software development life-cycle is studied in~\cite{suri2015object}.

Unlike the above-mentioned studies on software testability, which explore what makes software hard to test in general, this paper focuses on finding program features that define a CUT in such a way that can explain why the different testing techniques perform differently on it. The most closely related studies are Oliveira et al.~\cite{oliveira2018mapping} and Ferrer et al.~\cite{ferrer2010correlation}. The former is discussed in Section~\ref{sec:introduction} while later finding the correlation between static features of a program and techniques' performance. It uses Spearman’s rank correlation coefficient to find the correlation between eight static program features and coverage achieved on these programs. The CUTs used in the evaluation of the proposed technique are artificially generated with some specific values of the feature range within a predefined limit. The generated benchmark is then divided into sub-benchmarks, each having CUTs of a specific feature type. Test coverage achieved by techniques on each sub-benchmark is then used to find which technique performs better/worst in the presence of a particular feature. This work, however, cannot scale to a larger number of features or techniques. Furthermore, the artificially generated CUTs cannot represent the real software classes, which are characterised by a diverse set of features. The study also lacks in providing any insights into the impact of various features on the strength and weaknesses of a technique.

\section{Conclusion and Future Work} \label{sec:conclusion}

This paper presents the analysis of the strengths and weaknesses of various automated testing techniques using the Instance Space Analysis framework. A large collection of instances are compiled from software testing benchmark repositories, described in terms of meta-features and tested using 6 unit testing techniques. A subset of features is selected from the complete feature set, which is then used to generate a two-dimensional instance space for the visual analysis of the performance of selected techniques. The generated instance space shows that the instances from Quadrant 4, characterised by lower nesting depth, a small number of try/catch blocks, and fewer methods having $cc > 10$ are easy to be covered by all the techniques. On the contrary, instances in Quadrant 2 are harder to achieve good coverage by all the techniques due to their higher efferent coupling and large size (\textit{loc}). We also notice that the techniques having similar performance in terms of average coverage have different footprint areas and different strengths and weaknesses. Furthermore, it can be visualised that the benchmarks commonly used in the field of automated testing cover a smaller region of the complete instance space and are arguably insufficient to capture the full diversity of test instances required to comprehensively expose the strengths and weaknesses of the techniques.

The results shown in this paper are dependent on many factors including the chosen meta-data, parameter settings, the definition of good performance etc. Furthermore, the performance of the portfolio of techniques depends on their implementation and may also vary by changing the search budget. However, we use the Instance Space Analysis to gain visual insights when considering these settings, knowing that the methodology is general and repeatable, and in the expectation that the online instance space tool~\cite{matildagithub2020} and our sharing of meta-data~\cite{matilda2019} will facilitate the reproducibility and further exploration of this work. 

In the future, we plan to include more benchmarks, including various projects from defects4J~\cite{just2014defects4j}. Additional techniques can also be added, especially those which are not search based. Another interesting study would be to use mutation score as a performance metric. This would require to further enhance the feature space by adding more features that would make a CUT easy/hard for mutation tool. Lastly, we plan to analyse the performance of techniques using different search budgets, as search based techniques may perform differently given more time to explore the search space. 

\section*{Acknowledgements}

The funding for this research is provided by the Australian Research Council under grants DP210100041 and FL140100012.

\bibliography{mybibfile}
\clearpage

\begin{IEEEbiography}
    [{\includegraphics[width=1in,height=1.25in,clip,keepaspectratio]{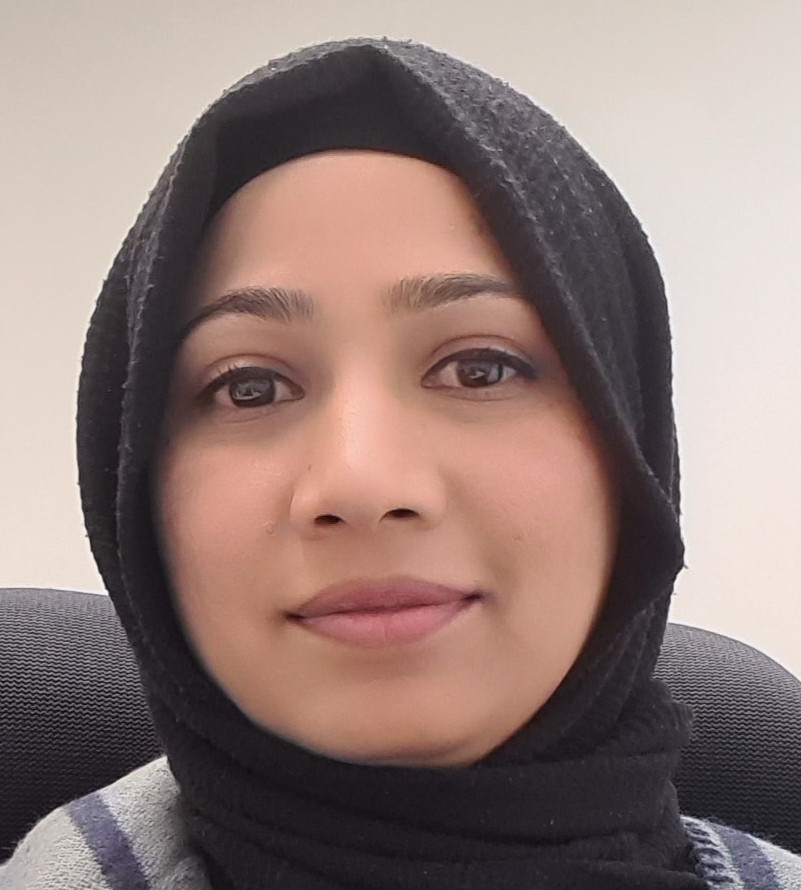}}]{Neelofar} received the bachelor’s and master’s degrees in Software Engineering from National University of Science and Technology, Pakistan, and the Ph.D. degree in Software Engineering from the University of Melbourne, in 2017. Currently, she is a research fellow at the Faculty of Information Technology, Monash University. 
    
    Her research interests include Search-Based Software Engineering (SBSE), Fault Localization and Automated Program Debugging. Her current work focuses on objective assessment of quality of Machine learning systems, in particular, Autonomous Vehicles (AVs). Neelofar is a member of program and organising committees of MSR2023 and EASE2023. 
\end{IEEEbiography}
\begin{IEEEbiography}
    [{\includegraphics[width=1in,height=1.25in,clip,keepaspectratio]{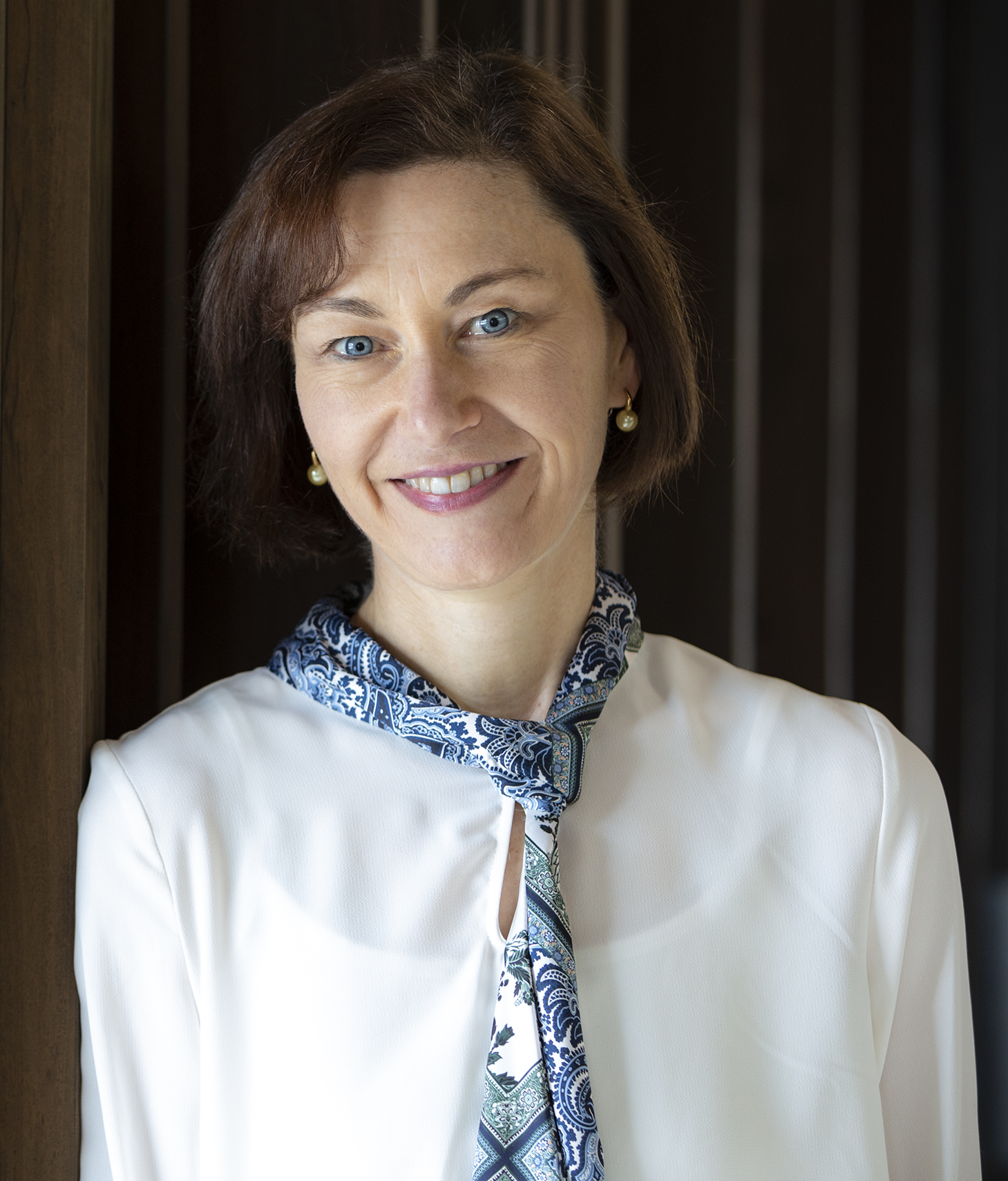}}]{Kate Smith-Miles (Senior Member, IEEE)} received the BSc (Hons) degree in mathematics in 1993, and the PhD degree in 1996, both from the University of Melbourne, Australia. She has held professorships in three disciplines: in Information Technology at Monash University, in Engineering at Deakin University, and in Mathematics at Monash University. She is currently a Melbourne Laureate Professor of applied mathematics in the School of Mathematics and Statistics at the University of Melbourne, and Associate Dean (Enterprise and Innovation) for the Faculty of Science. She was awarded an Australian Laureate Fellowship during 2014–2019 from the Australian Research Council, and has been honored with several medals for outstanding research contributions by the Australian Mathematical Society in 2010, the Australian and New Zealand Industrial and Applied Mathematics Society in 2017, and the Australian Society for Operations Research, in 2019. She has published around 300 refereed journal and international conference papers in the areas of neural networks, combinatorial optimization, machine learning, and applied mathematics. Her interdisciplinary collaborations have included applying mathematics to a wide range of problems from neuroscience to economic applications. She is an elected fellow of the Australian Mathematical Society, serving as president during 2016–2018, as well as a fellow of the Australian Academy of Science, and Engineers Australia.
\end{IEEEbiography}
\begin{IEEEbiography}
    [{\includegraphics[width=1in,height=1.25in,clip,keepaspectratio]{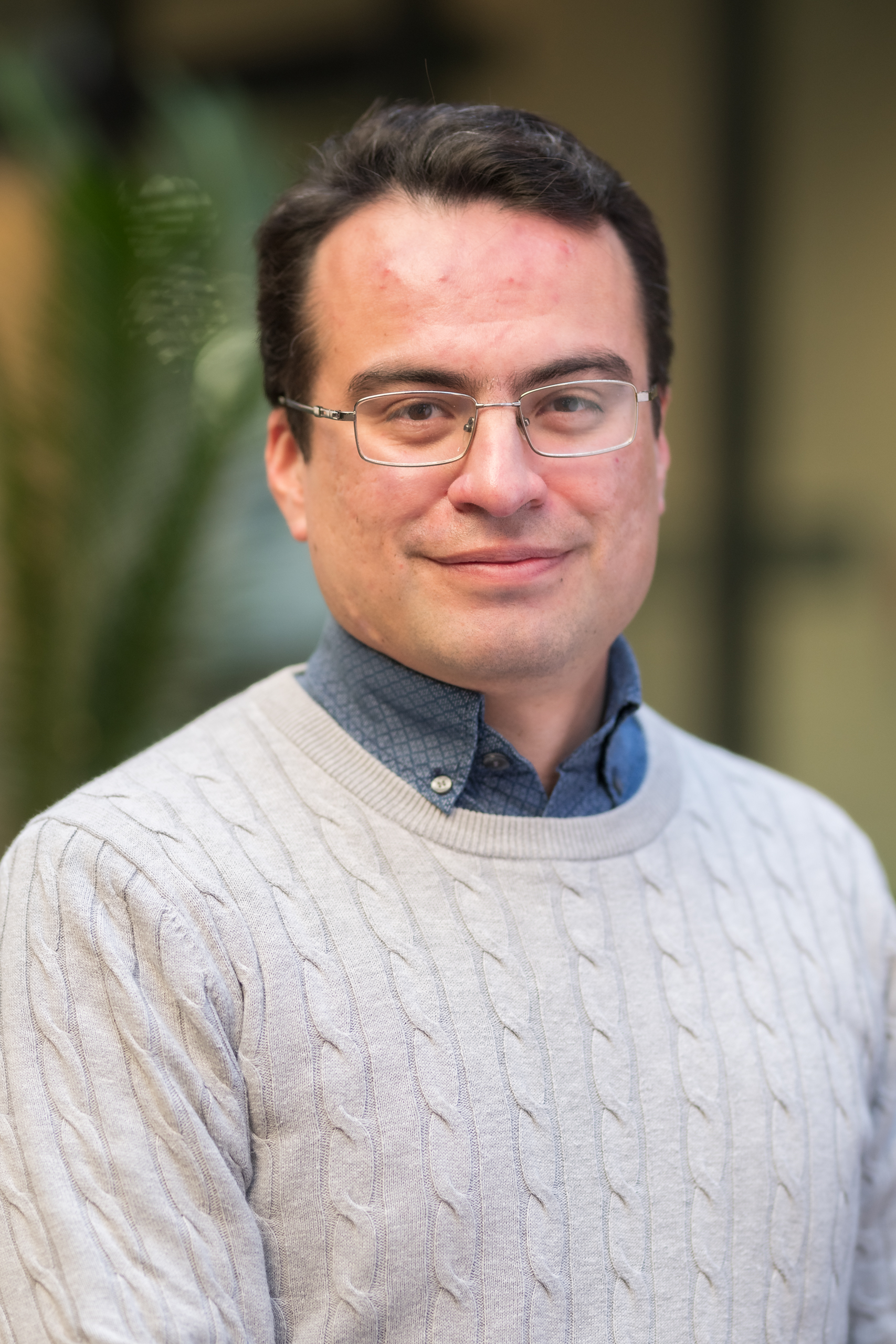}}]{Mario Andrés Muñoz} received the B.Eng. and M.Eng. degrees in Electronics Engineering from Universidad del Valle, Colombia, in 2005 and 2008 respectively, and the Ph.D. degree in Engineering from The University of Melbourne, Australia, in 2014. Currently, he is a Researcher at the School of Computer and Information Systems, The University of Melbourne; and the ARC Training Centre in Optimisation Technologies, Integrated Methodologies and Applications (OPTIMA). He has published over 50 papers.

His research interests focus on the application of optimisation, computational intelligence, signal processing, data analysis, and machine learning methods to ill-defined science, engineering and medicine problems
\end{IEEEbiography}
\begin{IEEEbiography}
    [{\includegraphics[width=1in,height=1.25in,clip,keepaspectratio]{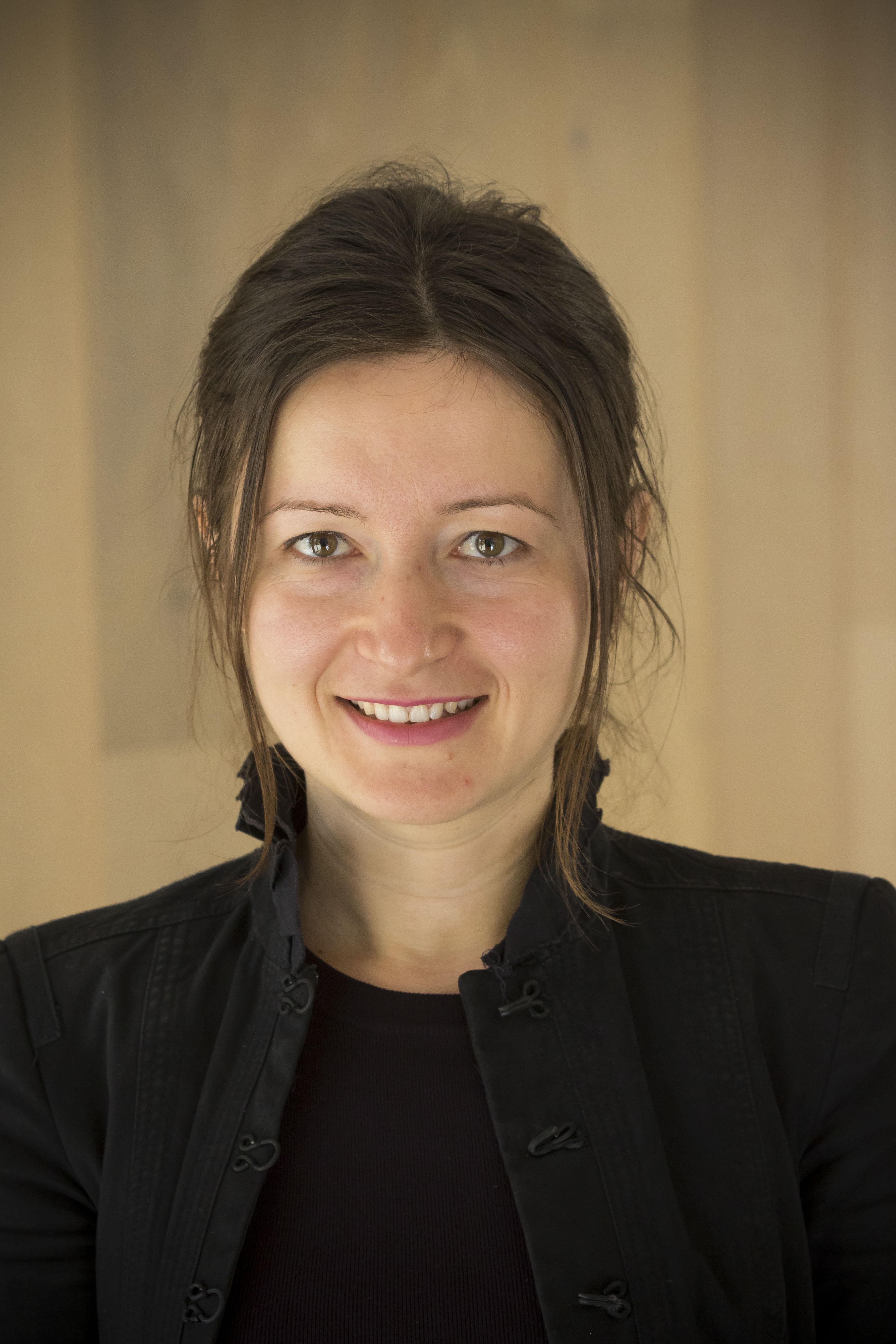}}]{Aldeida Aleti} is an Associate Professor and Associate Dean of Engagement and Impact at the Faculty of Information Technology, Monash University. Aldeida works in the area of Search-Based Software Engineering (SBSE), with a particular focus on the methodological aspects of how to assess the effectiveness of these techniques. This includes fitness landscape characterisation to analyse how hard optimisation problems are for search techniques, and algorithm selection, which is about identifying which optimisation technique works in what scenario. Aldeida has served in the program and organising committees of both SE and optimisation conferences, such as ASE, ICSE, GECCO, SSBSE, FSE, ICSA, etc., is a Senior Associate Editor for JSS, and Associate Editor for TOSEM and EMSE. Aldeida has attracted more than \$2.500,000 in competitive research funding and was awarded a Discovery Early Career Researcher (DECRA) Award from the Australian Research Council.
\end{IEEEbiography}

\end{document}